\documentclass[structabstract]{aa} 

\usepackage{graphicx}
\usepackage{txfonts}

\usepackage{natbib}
\bibpunct{(}{)}{;}{a}{}{,} 
\defcitealias{leger2009}{L{\'e}ger, Rouan, Schneider et al. 2009}
\defcitealias{borucki2011b}{B11b}
\usepackage{longtable,lscape}

\newcommand{\noNewCandidates}{43} 
\newcommand{\noNewCandidatesBatalha}{15}
\newcommand{\noNewSystems}{8}
\newcommand{\noNewSystemsBatalha}{3}

\newcommand{\noBoruckiFiltering}{14}

\newcommand{\frcn}[2]{ {\displaystyle \frac{#1}{#2} } }

\newcommand{\norm}[1]{||#1||}

\newcommand{\corot}{\textsl{CoRoT}}
\newcommand{\kepler}{\textsl{Kepler}}
\newcommand{\dst}{DST}

\begin{document}
%
\title{A study of the performance of the transit detection tool DST in space-based surveys}
\subtitle{Application of the CoRoT pipeline to Kepler data}

\author{
J.~Cabrera\inst{\ref{DLR}}
\and Sz.~Csizmadia\inst{\ref{DLR}} 
\and A.~Erikson\inst{\ref{DLR}}
\and H.~Rauer\inst{\ref{DLR},\ref{ZAA}} 
\and S.~Kirste\inst{\ref{DLR}}
}

\institute{
Institute of Planetary Research, German Aerospace Center, Rutherfordstrasse 2, 12489 Berlin, Germany\label{DLR}
\and Center for Astronomy and Astrophysics, TU Berlin, Hardenbergstr. 36, 10623 Berlin, Germany\label{ZAA}
}

\date{Received ; accepted }

\abstract
{
  Transit detection algorithms are mathematical tools used for
  detecting planets in the photometric data of transit surveys.
  In this work we study their application to space-based surveys.
}
{
  Space missions are exploring the parameter space of the transit 
  surveys where classical algorithms do not perform optimally, either
  because of the challenging signal-to-noise ratio of the signal or
  its non-periodic characteristics.
  We have developed an algorithm addressing these challenges for the
  mission \corot.
  Here we extend the application to the data from the space mission
  Kepler.
  We aim at understanding the performances of algorithms in different
  data sets. 
}
{
  We built a simple analytical model of the transit signal and
  developed a strategy for the search that improves the detection 
  performance for transiting planets.
  We analyzed Kepler data with a set of stellar activity filtering
  and transit detection tools from the \corot\,community that are
  designed for the search of transiting planets.
}
{
  We present a new algorithm and its performances compared to one of
  the most widely used techniques in the literature using
  \corot\,data. 
  Additionally, we analyzed Kepler data corresponding to quarter Q1
  and compare our results with the most recent list of planetary 
  candidates from the Kepler survey.
  We found candidates that went unnoticed by the Kepler team when
  analyzing longer data sets. 
  We study the impact of instrumental features on the production of
  false alarms and false positives.
  These results show that the analysis of space mission data advocates 
  the use of complementary detrending and transit detection tools
  also for future space-based transit surveys such as PLATO.
}
{}

\keywords{
  Methods: data analysis; 
  Techniques: photometric; 
  Stars: planetary systems;
  Stars: binaries: eclipsing
}

\maketitle

%
\section{Introduction}

Transit detection algorithms are mathematical tools that aim at
detecting the signal of a transiting planet in the photometric time
series of a star (light curve).
The generally accepted characteristics of the planet transit are a
short, small, periodic decrease of the (assumed) constant luminosity
flux of the star caused by the interposition of the opaque planet in
front of the stellar disk, as seen from the point of view of the
observer.
Transits are short because the duration of a transit is roughly
proportional to the quotient of the radius of the star over the
semi-major axis of the orbit of the planet  ($t_T/P \sim R_*/\pi a;$
see \citealt{seager2003}). 
This means that even for the shortest known period exoplanets orbiting 
at 3 or 4 stellar radii, such as 
55 Cnc e \citep{winn2011a}, 
WASP-19b \citep{hebb2010}, 
WASP-43b \citep{hellier2011},
Kepler-10b \citep{batalha2011}, 
CoRoT-7b \citepalias{leger2009},
or WASP-18b \citep{hellier2009} 
(all of them with orbital periods shorter than 24h) the duration is
shorter than one tenth of the orbital period. 
Transit depths are small because the diminution of the flux is
proportional to the square of the quotient of the radii of the planet
and the star: $\Delta F/F \sim \left(R_p/R_*\right)^2$. 

The arrival of the space-based surveys of exoplanets such as \corot\,
\citep{baglin2006} and \kepler\,\citep{borucki2010a} has placed
transit surveys in the region of the parameter space where transits
are not short, in terms of hours, any more.
For example, the transit of CoRoT-9b \citep{deeg2010} lasts 8h, but
there are also detections of giant eclipsing binaries where the
eclipses last several days, like CoRoT 101126445, whose eclipses last
49h for a period of 30 days \citep{cabrera2009}. 
At the same time, the detected signals are not weak either.
In this case there are two tendencies, on the one hand toward really 
shallow transits, such as those of the terrestrial planets 55 Cnc e,
CoRoT-7b or Kepler-10b, and on the other hand toward transits of small
planets around small stars, which enables detecting Earth-sized
planets from the ground, such as GJ-1214b \citep{charbonneau2009}.
Finally there are detections of transits that are not periodic
anymore, the transit timing variations of Kepler-9c \citep{holman2010}
reach an amplitude of 140 minutes, comparable with the length of the
transit, in a time span of 200 days.

Space surveys of transiting planets such as \corot\,and \kepler\,have
improved our knowledge about the performances of transit detection
techniques.
Between 2004 and 2007 the \corot\, community made a series of studies to
test the transit detection capabilities of different algorithms. 
The analysis of the results of these tests, published by
\citet{moutou2005,moutou2007}, revealed on the one hand that all
algorithms had competitive performances and yielded comparable
results, but on the other hand it was shown that both false alarms and
non-detections were method dependent. 
The transit detection approach followed by the \corot\, community is
based on the results of these analyses: combine the results of
different teams using different detrending and detection techniques to
reduce the rate of false alarms and to push the detection limit as
far as possible.

The \kepler\,survey has followed a different approach using one single
detrending and detection algorithm \citep{jenkins2002a,jenkins2010c}
whose performance has been extensively studied and tested
\citep{jenkins2002b,jenkins2010a}. 
In view of future transit search missions, e.g. PLATO
\citep{catala2009}, it is interesting to compare the different
approaches.

There are several tools in the literature dealing with the automatic
detection of transiting candidates 
\citep{aigrain2002,bonomo2008,borde2007,carpano2008,colliercameron2006,
defay2001,grziwa2012,jenkins1996,jenkins2002a,kovacs2002,ofir2008,
protopapas2005,regulo2007,renner2008,schwarzenberg-czerny2006}.  
In addition to those, \citet{schwarzenberg-czerny1998} gives a very 
interesting discussion on the general techniques for the detection of
periodic signals. 

The code presented in this paper has been used regularly since 2006 in
the \corot\, survey for transiting planets
\citep{carpano2009,cabrera2009,erikson2012,carone2012,cavarroc2012}.
The algorithm, called \dst, 
({\em D{\'e}tection Sp{\'e}cialis{\'e}e de Transits}) addresses some
of the challenges that face current and future transit surveys.
This algorithm aims at a specialized detection of transits by
improving the consideration of the transit shape and the presence of
transit timing variations.
As a first step, we have applied our \corot\, detrending and transit
detection tools to the public data release Q1 \citep{borucki2011a} and
compared the results to the list of planetary candidates published by
\kepler\, (\citealt{borucki2011b}, hereafter \citetalias{borucki2011b},
and \citealt{batalha2012}).
We aim at testing the capabilities of a different algorithm to find
new transit candidates in a reduced data set, because we also aim at
understanding the dependence of the yield as a function of the length
of the observations. 
Such a study was already started in CoRoT \citep{cabrera2009} and
it is of great interest when deciding the observation strategy of
transit surveys, for example when deciding the optimimum length of an
observational run.
On the other hand, it is important to find successful algorithms that
are able to detect planets in data sets with a reduced number of
transit events. 
This is the case of modern transit surveys aiming at detecting
planetary candidates with periods comparable to the total length of
the observations, for example planets in the habitable zone of
solar-like stars. 

Section~\ref{sec:ecuaciones} contains the mathematical description of
the algorithm.
Section~\ref{sec:discusion} compares the performance of the new
algorithm with a widely used technique, the BLS code
\citep{kovacs2002}.
We do not aim to reproduce the algorithm comparisons found in the
literature such as the analysis made by
\citet{tingley2003a,tingley2003b} or \citet{moutou2005,moutou2007} 
but to present the differences of this new code with respect to one of
the most widely used codes (BLS) and explain its advantages.
Section~\ref{sec:datos} describes the treatment applied to \kepler\,Q1
data. 
Section \ref{sec:resultados} describes the results of the test,
including a number of new transiting candidates. 
Finally, Section~\ref{sec:summary} summarizes the outcome and
describes future developments. 

%
\section{The transit search algorithm}
\label{sec:ecuaciones}

A light curve is a time series of the photometric observations of a
star. 
Mathematically, it can be seen as a collection of pairs of time-flux,
which can be translated into a function like
\begin{equation}
  \label{eq:light_curve}
  g(x) = \sum_{i=0}^{N-1} g_i \, \Pi_h(x-x_i);
\end{equation}
where $(x_i,g_i)$ are the $N$ pairs time-flux and
\begin{equation}
  \label{eq:step_function}
  \Pi_h(x-x_i) = \left\{
  \begin{array}{ll}
    1, & x_i \le x < x_i+h; \\
    0, & \mathrm{elsewhere},
  \end{array}
  \right.
\end{equation}
with $h$ being the integration time. 
$g(x)$ should not be understood as the function representing the
continuous time waveform corresponding to the photometric
measurement, but as a mathematical tool to represent photometric
observations. 
If $G(t)$ represents the continuous waveform coming from the star,
$g_i$ is defined as
\begin{equation}
  \label{eq:def_of_g}
  g_i = \frac{1}{h} \int_{x_i}^{x_i+h} G(t) dt.
\end{equation}

The BLS \citep{kovacs2002} is arguably the most popular transit
detection algorithm in the literature.
It proposes a box model for the transit shape, which can be described
as
\begin{equation}
  \label{eq:function_BLS}
  f(x) = \left\{
  \begin{array}{ll}
    L, & x_{i} \in T \\
    H, & \mathrm{elsewhere},
  \end{array}
\right.
\end{equation}
where $H$ is the mean value of the flux out of transit and $H-L$ is
the depth of the transit, which occurs in the interval $T$. 
The test statistic defined by BLS is the distance between the
functions $f$ and $g$, defined as
\begin{equation}
  \label{eq:BLS_statistic_definicion}
  \mathcal{D}^2 = \int \omega \left( f - g \right)^2,
\end{equation}
where we have defined the weight function analogously to $g$
\begin{equation}
  \label{eq:weight}
  \omega(x) = \sum_{i=0}^{N-1} \omega_i \, \Pi_h(x-x_i),
\end{equation}
keeping the definition of \citet{kovacs2002}: 
$\omega_i = \sigma_i^{-2} \left[ \sum_{j=0}^{N-1} \sigma_j^{-2} \right]^{-1}$.
It is trivial to prove that (\ref{eq:BLS_statistic_definicion}) is the
product of the constant $h$ (the integration time) times the test 
statistic defined by equation 1 in \citet{kovacs2002}. 
Expression (\ref{eq:BLS_statistic_definicion}) can be minimized 
analytically with respect to the variables $H$ and $L$ once the
transit region $T$ is known:
\begin{equation}
  \label{eq:BLS_H_L}
  \renewcommand{\arraystretch}{2.5}
  \begin{array}{r@{\,=\,}l}
    H & \frcn{ \displaystyle \sum_{i=0}^{N-1} \omega_i g_i - \sum_{ i \in T} \omega_i g_i}{ \displaystyle \sum_{i=0}^{N-1} \omega_i - \sum_{ i \in T} \omega_i }, \\
    L & \frcn{ \displaystyle \sum_{ i \in T} \omega_i g_i}{ \displaystyle \sum_{ i \in T} \omega_i },
  \end{array}
\end{equation}
and the BLS test statistic is
\begin{equation}
  \label{eq:BLS_statistic_final}
  \mathcal{D}^2 = h \left[ \sum_{i=0}^{N-1} \omega_i g_i^2 - 
    \frcn{ \left( \displaystyle \sum_{i=0}^{N-1} \omega_i g_i - \sum_{i \in T} \omega_i g_i \right)^2 }{ \displaystyle \sum_{i=0}^{N-1} \omega_i - \sum_{i \in T} \omega_i } - 
   \frcn{ \left( \displaystyle \sum_{i \in T} \omega_i g_i \right)^2 }{ \displaystyle \sum_{i \in T} \omega_i } \right],
\end{equation}
which is further simplified by making
$\sum_{i=0}^{N-1} \omega_i g_i = 0,$ and
$\sum_{i=0}^{N-1} \omega_i = 1;$ this finally produces equation 4 in
\citet{kovacs2002}.

In practice, as the ephemeris of the transit (period $P$,
epoch $x_0$ and transit length $d$) are not known a priori, one folds
and bins the light curve to a test period and then calculates the test 
statistic for every possible value of $x_0$ and $d$. 
The minimum of the test statistic corresponds to the ephemeris of the
transit. 
If there is no significant extreme value of the test statistic, the
conclusion is that there is no detectable transit in the light curve
at the tested period. 

We can define an alternative model to the transit function using a
parabola instead of a box function and modifying the definition of the
region in transit. 
The advantages of these choices are shown in
Section~\ref{sec:discusion}. 
The model function will be
\begin{equation}
  \label{eq:function_Dst}
  f(x) = \left\{
  \begin{array}{ll}
    H - \delta_k + \frcn{4 \delta_k}{d_k^2} \left(x - x_{ok}\right)^2, & x \in T_k; \; \\
    H, & x \notin T_k.
  \end{array}
  \right.
\end{equation}
We assume that there are K observed transits, each transit
denoted as $T_k$ centered in the position $x_{ok}$. The definition
of the transits is discussed in the next section.
The duration of the $k$th transit $T_k$ is given by $d_k$ and its
depth by $\delta_k$. 
The definition of the distance between the functions $f$ and $g$ is
identical to (\ref{eq:BLS_statistic_definicion}) and its minimization
with respect to the variables $H$ and $\delta_k$ yields the following 
constrain: 
\begin{equation} \label{eq:Dst_H_delta}
  \renewcommand{\arraystretch}{2.5}
  \begin{array}{r@{\,=\,}l}
    H & \frcn{ \displaystyle \sum_{i=0}^{N-1} \omega_i g_i - \sum_{k} \frcn{\Lambda_k^2 g\Lambda_k^2}{\Lambda_k^4} }{ \displaystyle \sum_{i=0}^{N-1} \omega_i - \sum_{k} \frcn{\left(\Lambda_k^2\right)^2}{\Lambda_k^4} }, \\
    \delta_k & \frcn{d_k^2}{4} \frcn{ g\Lambda_k^2 - H \Lambda_k^2}{\Lambda_k^4},
  \end{array}
\end{equation}
where
\begin{equation} \label{eq:Dst_Lambdas}
  \renewcommand{\arraystretch}{2.5}
  \begin{array}{r@{\,}l}
    \Lambda_k^2  & \displaystyle = \sum_{i \in T_k} \omega_i \left[ \left( x_i - x_{ok} + \frcn{h}{2} \right)^2 - \frcn{d_k^2}{4} + \frcn{h^2}{12} \right], \\
    g\Lambda_k^2 & \displaystyle = \sum_{i \in T_k} \omega_i \left[ \left( x_i - x_{ok} + \frcn{h}{2} \right)^2 - \frcn{d_k^2}{4} + \frcn{h^2}{12} \right] g_i, \\
    \Lambda_k^4  & \displaystyle = \sum_{i \in T_k} \omega_i \left[ \left( x_i - x_{ok} + \frcn{h}{2} \right)^4 + \frcn{1}{2} \left( h^2 - d^2 \right) \left( x_i - x_{ok} - \frcn{h}{2} \right)^2 \right. \\
                 & \displaystyle \qquad \left.+ \frcn{1}{8} \left( \frcn{h^4}{10} - \frcn{d_k^2 h^2}{3} + \frcn{d_k^4}{2} \right) \right].
  \end{array}
\end{equation}
Using the above constrains, the distance between the functions $f$ and
$g$ becomes
\begin{equation} \label{eq:Dst_statistic}
\mathcal{D}^2 = h \left[ \sum_{i=0}^{N-1} \omega_i g_i^2 - H \sum_{i=0}^{N-1} \omega_i g_i - \sum_{k} \frcn{4 \delta_k}{d_k^2} g\Lambda_k^2 \right].
\end{equation}

However, contrary to \citet{kovacs2002}, we are not going to define the
test statistic as the distance between the functions $f$ and $g$ (or,
in the case of BLS, the signal residue resulting from subtracting the
constant term $\sum_{i=0}^{N-1} \omega_i g_i^2$) from
(\ref{eq:BLS_statistic_final}).
Following \citet{schwarzenberg-czerny1998}, for a signal $x$ that is
decomposed into its model $x_{\parallel}$ and residuals 
$x_{\perp} = x - x_{\parallel},$ the statistic used to assess if the
model is an appropriate description of the data will be 
$\norm{x_{\parallel}}^2 / \norm{x_{\perp}}^2$, where $\norm{x}^2=\sum x_i^2$.
As indicated in the previously cited paper, we can draw the same 
conclusions with other statistics provided that they are uniquely 
related (see also \citealt{jenkins1996,tingley2003b,schwarzenberg-czerny2006}).
In our notation, $x=g$ and $x_{\parallel}=f$, whereas 
$\mathcal{D} = \norm{f-g} = \norm{ x - x_{\parallel} } = \norm{x_{\perp}};$ 
therefore the statistic is 
$\norm{x_{\parallel}}^2/\norm{x_{\perp}}^2 = \norm{f}^2/\mathcal{D}^2$. 
In this case
\begin{equation} 
  \label{eq:Dst_normf}
  \norm{f}^2 = h \left( \sum_{i=0}^{N-1} \omega_i g_i + \sum_{k} \frcn{4 \delta_k}{d^2} g\Lambda_k^2 \right),
\end{equation}
and finally the test statistic is
\begin{equation}
\label{eq:Dst_estimador}
\frcn{\norm{f}^2}{\mathcal{D}^2} = 
  \frac{ \displaystyle \sum_{i=0}^{N-1} \omega_i g_i + \sum_{k} \frcn{4 \delta_k}{d_k^2} g\Lambda_k^2}
       { \displaystyle \sum_{i=0}^{N-1} \omega_i g_i^2 - H \sum_{i=0}^{N-1} \omega_i g_i - \sum_{k} \frcn{4 \delta_k}{d_k^2} g\Lambda_k^2 }.
\end{equation}

%
\section{Discussion of the transit search algorithm}
\label{sec:discusion}

The test statistic defined by (\ref{eq:Dst_estimador}) is more
complicated than defined by BLS (see \ref{eq:BLS_statistic_final}) and
hence it involves a heavier computational effort. 
However, it has four main advantages:
1) it provides a better description of the transit shape, which in
turn provides a better behavior of the algorithm in the presence of  
transits; 
2) it uses the same number of free parameters as BLS; it is always
possible to improve the model of the transit shape by using more free
parameters (for example, a trapezoid has one more free parameter: the
duration of the ingress phase), but at the expense of computational
effort;
3) 
our estimator performs better than BLS for the reasons explained
below,
and 4) the definition of the region in transit is more flexible. 

To illustrate these assertions, we have analyzed public data from the 
satellite CoRoT\footnote{CoRoT public data is available thorough the
  CoRoT Data Archive hosted by IAS: 
  {\tt http://idoc-corot.ias.u-psud.fr.}} with both BLS and the new 
method described in this paper (\dst) and compared their performance
in two test cases. 
In the first one (Subsections~\ref{subsec:shape} and
\ref{subsec:statistic}), we show how the \dst\, model of the transit
shape improves the transit detection efficiency.
We have chosen two typical targets of transit surveys for this test, 
a planet, CoRoT-1b \citep{barge2008a}, and an eclipsing binary, CoRoT
102763847 \citep{carpano2009}, of similar depths and durations, 
although different periods.
We chose an eclipsing binary because although transit surveys
aim at detecting planets, they find many eclipsing binaries as 
a by-product.
The probability to detect eclipsing binaries is relatively higher
because they are numerous and because the S/N ratio of eclipses is in 
general stronger than the S/N ratio of a planetary transit, since in
stellar binary systems both members have in general comparable sizes. 
By choosing a planet and an eclipsing binary of similar depths and
brightness we can compare the S/N ratio of the detection in typical
targets of planet detection surveys.
Both targets were observed at the CoRoT sampling 
rates\footnote{CoRoT-1b was observed at 512s and 32s sampling rate
  during the run IRa01, described in \citet{carpano2009}, where it got
  the win\_id IRa01\_E2\_1126. The eclipsing binary with CoRoT id
  102763847 was observed at 32s sampling rate during the run IRa01,
  where it got the win\_id IRa01\_E1\_1158, and at 512s sampling rate
  during the run LRa01 \citep{carone2012}, where it got the win\_id
  LRa01\_E2\_0588.} 
of 32s and 512s and the S/N ratio achieved are representative of
typical giant planet signals in transit surveys.
This first test case consists of the analysis of a section of the light
curve containing only one transit, respectively one eclipse.
We implemented the BLS and \dst\, algorithms\footnote{the code
  is available on request to the author.} and we applied
them on the same data set to compare their performances. 

But in addition to Jupiter planets, transit surveys also aim at
detecting small terrestrial planets.
In our second test case (Subsection~\ref{subsec:c7b}) we compare the 
detectability with BLS and \dst\,of CoRoT-7b, representative of the
dim signals of Earth-sized planets.
In this case we performed a blind search for the period of the
transiting planet using the same filtered light curve for both
algorithms. 

\subsection{Test case I: the transit shape}
\label{subsec:shape}

A second-order polynomial is expected to yield a better description of
the transit shape than a box model because it includes a description,
although incomplete, of the ingress and egress of the transit, which
the box-shaped description simply ignores. 
The number of free parameters in \dst\, and BLS is the same: the value
of the flux out of transit $H$, the depth of the transit $\delta_k = H
- L$, and the ephemeris of the transit (period $P$, epoch $x_0$, and
duration $d$). 
This is in contrast to the trapezoidal model of the transit, which 
involves one more free parameter (the duration of the ingress phase).
As we are analyzing the data of a single event, we fix the value of
the period in the following analysis .

\begin{figure}[t]
  \begin{center}
    \includegraphics[%
      width=0.9\linewidth,%
      height=0.5\textheight,%
      keepaspectratio]{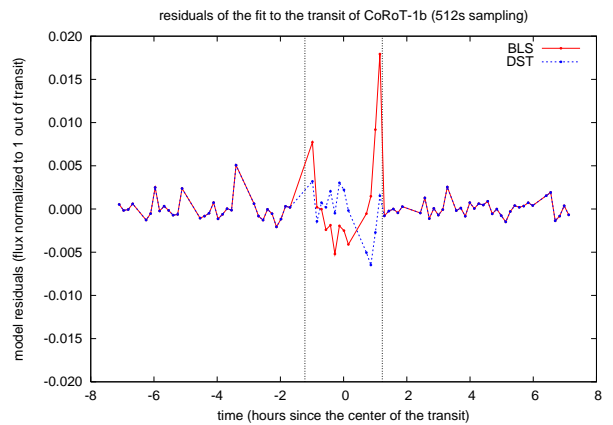}
  \end{center}
  \caption{
    Residuals of the BLS and \dst\,model of CoRoT-1b sampled at
    512s. The central time of the transit chosen has a Julian Date
    value of $2\,454\,142.8547$ days. The dotted vertical lines
    indicate the duration of the transit.
  }
  \label{fig:c1b512}
\end{figure}

\begin{figure}[t]
  \begin{center}
    \includegraphics[%
      width=0.9\linewidth,%
      height=0.5\textheight,%
      keepaspectratio]{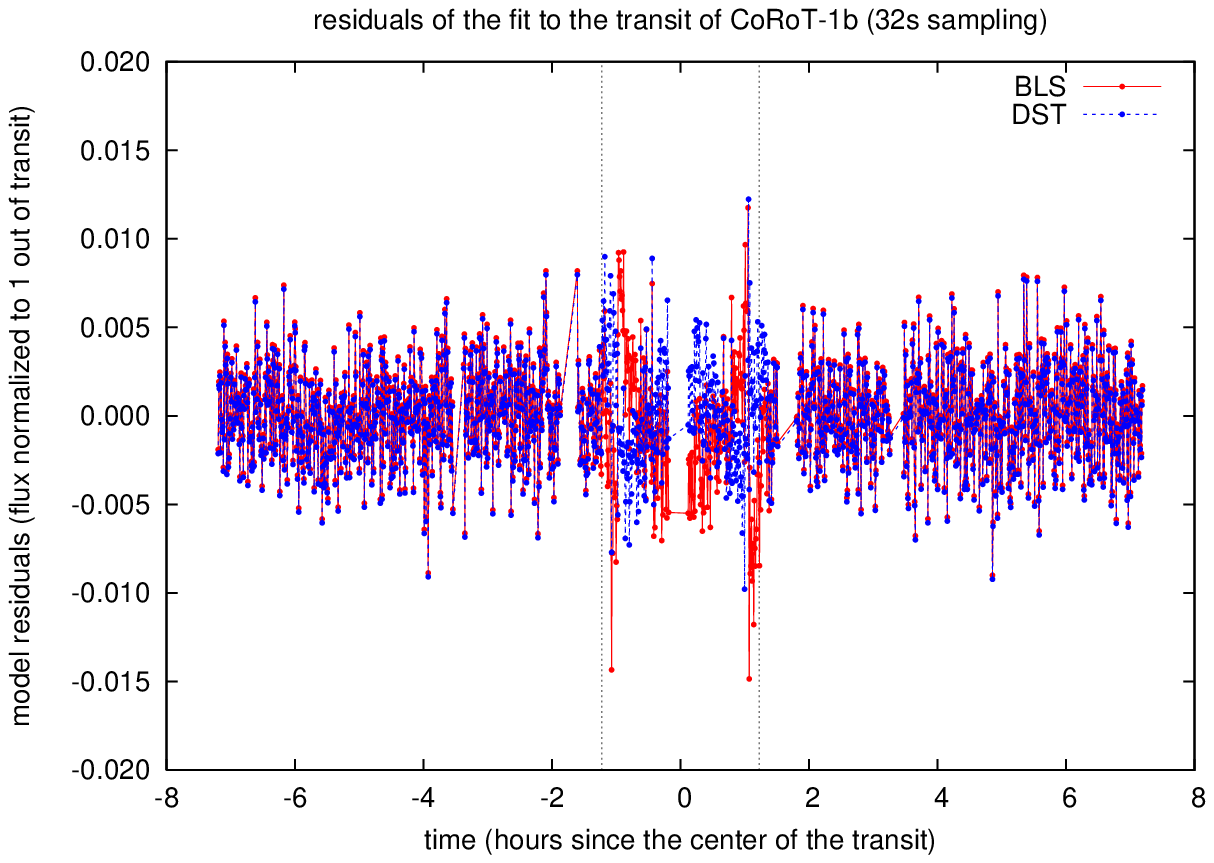}
  \end{center}
  \caption{
    Residuals of the BLS and \dst\,model of CoRoT-1b sampled at
    32s. The central time of the transit chosen has a Julian Date
    value of $2\,454\,191.1413$ days. The dotted vertical lines
    indicate the duration of the transit.
  }
  \label{fig:c1b32}
\end{figure}

We used the BLS and the \dst\, algorithms to fit the aforementioned
free parameters around the position of the chosen events. 
We show in Fig.~\ref{fig:c1b512} the residuals of the fit of the
CoRoT-1b transit sampled at 512s with BLS and \dst. 
Figure~\ref{fig:c1b32} shows the same residuals for the 32s
sampling. 
These figures show that BLS fails to reproduce the transit shape in
the ingress and egress, as expected. 
In Section~\ref{sec:ecuaciones} we have shown that the statistic used
by BLS and \dst\, is the distance between the observed data points and
the model function as described by expression
\ref{eq:BLS_statistic_definicion}. 
Table~\ref{table:residuals} shows that \dst\, is indeed a better
description of the transit and eclipse signal. 
When the S/N ratio is high (in the 512s sampled ratio), \dst\,
produces 68\% and 59\% less residuals than BLS (for the planet and the
eclipsing binary, respectively). 
When the S/N ratio is lower (in the 32s sampled ratio) the difference
is less accentuated and \dst\, performs around 14\% better in both 
targets. 

\begin{figure}[t]
  \begin{center}
    \includegraphics[%
      width=0.9\linewidth,%
      height=0.5\textheight,%
      keepaspectratio]{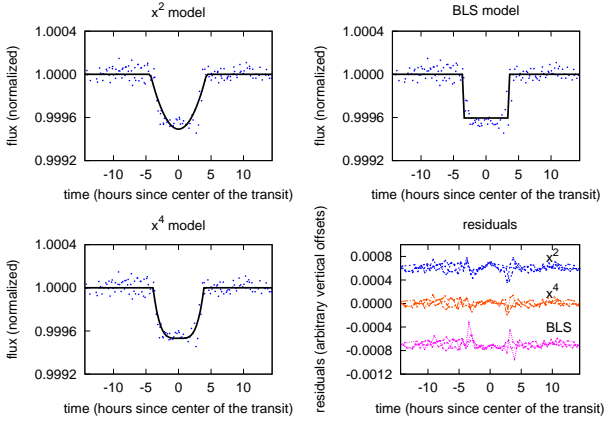}
  \end{center}
  \caption{
    Different models ($x^2,$ $x^4$ and BLS) for the light curve of
    Kepler-22b and comparison of the residuals.
  }
  \label{fig:k22b}
\end{figure}

\begin{figure}[t]
  \begin{center}
    \includegraphics[%
      width=0.9\linewidth,%
      height=0.5\textheight,%
      keepaspectratio]{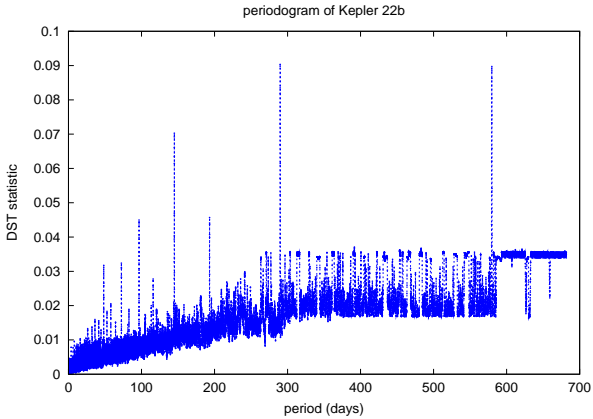}
  \end{center}
  \caption{
    Periodogram of the DST statistic for the light curve of
    Kepler-22b. This analysis includes 682 days of data from quarters
    Q0 to Q6 and there are three transit events.
  }
  \label{fig:k22b_pdg}
\end{figure}

It might be rightfully argued that the parabolic is not the best
description of the {\em flat bottom} feature of planetary transits. 
Indeed, there are functions that better reproduce the shape of long
transits, where the contribution of the {\em flat part} is most
relevant, such as the cases of the already discussed CoRoT-9b or, for
an example of a super-Earth, Kepler-22b \citep{borucki2012}.
Figure~\ref{fig:k22b} shows the binned light curve of Kepler-22b
modeled with the parabolic shape described in this paper (labeled
$x^2$), with a quadratic shape (labeled $x^4$) described by equation
\ref{eq:function_Dst_x4} below, and with BLS. 
Indeed, Kepler-22b, a $\sim 2$ Earth radii planet with an orbital
period of 290 days is not a challenging detection for DST considering
the distinct signal produced by the planet in the periodogram (see
Fig.~\ref{fig:k22b_pdg}).

\begin{equation}
  \label{eq:function_Dst_x4}
  f'(x) = \left\{
  \begin{array}{ll}
    H - \delta_k + \frcn{16 \delta_k}{d_k^4} \left(x - x_{ok}\right)^4, & x \in T_k; \; \\
    H, & x \notin T_k.
  \end{array}
  \right.
\end{equation}

If we compare the residuals of these models, we immediately see that
most of the residuals of BLS come from the ingress and egress phases,
as was the case with CoRoT-1b above.
We can also verify that the $x^4$ model performs slightly better than
the $x^2$ model, but we do not obtain the same degree of improvement as
when we compare $x^2$ and BLS.
Indeed, if $f(x)$ is the best function describing the shape of a
transit, and if this function is even (which is best true for circular
orbits and spherical stars), then the $x^2$ model is the first non-null
term of the Taylor series of the optimal function $f(x)$.
In some cases, the term in $x^4$ might be more representative, but it
will always be a more complicated description 
and it will be far from optimal for transits with higher impact parameters.
The parabolic model is not the best model, but it is the most
simple (with fewer free parameters), analytic, continuous (in
contrast to BLS) description of the transit. 

\begin{table}
  \caption{\label{table:residuals}Relative residuals of the BLS and \dst\,models of the CoRoT-1b and CoRoT id 102763847 targets in parts per million.} 
  \renewcommand{\arraystretch}{1.2}
  \begin{tabular}{lcc}
    \hline\hline
    target & $\int \left( f - g_\mathrm{BLS}\right)^2/\int f^2$ & $\int \left( f - g_\mathrm{DST}\right)^2/\int f^2$ \\
    \hline
    CoRoT-1b (512s)  & 7.7$\cdot 10^{-6}$ & 2.4$\cdot 10^{-6}$ \\
    CoRoT-1b (32s)   & 8.7$\cdot 10^{-6}$ & 7.5$\cdot 10^{-6}$ \\
    102763847 (512s) & 2.1$\cdot 10^{-6}$ & 0.8$\cdot 10^{-6}$ \\
    102763847 (32s)  & 7.2$\cdot 10^{-6}$ & 6.3$\cdot 10^{-6}$ \\
    \hline
  \end{tabular}
\end{table}

\subsection{Test case I: the test statistic for a giant planet}
\label{subsec:statistic}

It is difficult to compare the result of the test statistics of BLS
and \dst\, because they are defined in different ways and 
represent different magnitudes. 
The signal residue (SR) defined in \citet{kovacs2002} is derived from
expression (\ref{eq:BLS_statistic_definicion}), as has been shown
in Section~\ref{sec:ecuaciones}. 
For example, in our particular analysis of the light curve of
CoRoT-1b, the value of the SR, defined as $\sqrt{s^2/r(1-r)}$ in
\citet{kovacs2002}, at the ephemeris of the transit has a value of
$1.985\,18\cdot 10^6$, whereas it has a value of $1.985\,14\cdot 10^6$
outside the transit, which is a difference of about 22 ppm.
However, the statistic used in BLS is the signal detection efficiency
(SDE), built from the SR, which in turn has an on-transit value of
$2.5$, and this is the value used to determine the reliability of a
transit candidate.
The \dst\, uses expression (\ref{eq:Dst_estimador}), which is not
directly comparable to the SR of BLS. 
However, in the case of CoRoT-1b, the value of this estimator on
transit is 800 times higher than the value off transit 
(3.5 vs. 0.004).  
If we proceed as in \citet{kovacs2002} and define an equivalent to
the SDE for the \dst\,estimator (see Fig.~\ref{fig:c7b}),
its value on transit is 6.4, more than doubling the performance of BLS.  

One of the disadvantages of the SDE statistic is that it includes the
points in transit in the calculation of the reference standard
deviation used to assess the significance of a signal. 
This is a known fact already discussed by different authors
\citep[see, for example,][]{kovacs2002,schwarzenberg-czerny2006,jenkins2010c}.  
For example, the WASP survey uses the BLS code for transit detection,
but does not use the SDE as statistic to select planetary candidates,
but a different statistic with dimensions of $\chi^2$
\citep[see][]{colliercameron2006}. 



\subsection{Test case II: the test statistic for a terrestrial planet}
\label{subsec:c7b}

The difficulty of comparing the performance of different detection
algorithms rests in the definition of the test data set and in the
optimization of the codes. 
Simulated data sets have the advantage of handling the impact of the
S/N ratio in the detectability of the signals. 
However, in practice all transit surveys suffer from correlated noise
\citep{pont2006} that dominates the detectability of dim signals and
the rate of false alarms.
Therefore we chose for our analysis the measured light curve of
CoRoT-7b as test data set. 
We used standard techniques for iltering stellar variability and
instrumental residuals (see below).
We did not control how much the red noise affected the detectability
of the signal, but we assumed that it would be representative of the 
typical impact of this kind of perturbation in transit surveys.  
Lastly, a fair comparison of different techniques requires that the
configuration parameters used for the transit search algorithms are
optimized with the same rigor. 
We have done our best to put both algorithms, BLS and \dst, in the
most similar conditions.
Considering the results of the previous section,
we assume that the differences in the performance of the analysis of
CoRoT-7b that we present are indeed due to the advantages that \dst\,
entails over BLS and not due to an improper configuration of the
optimizing parameters.

\begin{figure}[t]
  \begin{center}
    \includegraphics[%
      width=0.9\linewidth,%
      height=0.5\textheight,%
      keepaspectratio]{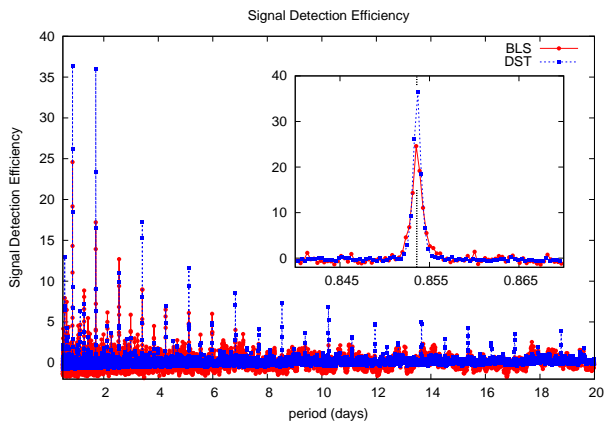}
  \end{center}
  \caption{
    Value of the signal detection efficiency SDE of BLS and \dst\,
    for the light curve of CoRoT-7b. Detail of the region around the
    measured orbital period of this planet, indicated with a vertical
    line  ($0.853\,585 \pm 0.000\,024$ days).
  }
  \label{fig:c7b}
\end{figure}

Figure~\ref{fig:c7b} shows the comparison of the SDE defined in the
previous subsection for BLS and \dst\,in the analysis of the same
light curve of CoRoT-7b. 
First, \dst\,provides a more robust detection, identifying
correctly the right period and many harmonics up to periods of 20 days,
and providing a less noisy result than BLS.
Second, \dst\,has an SDE of 35 compared to an SDE of only 24
for BLS.
This test case shows that \dst\, performs better than BLS also
detecting the dim signals of terrestrial planets. 
Interestingly, the SDE for CoRoT-7b (35) is bigger than the
SDE of CoRoT-1b (6.4).
There is no reason to be surprised. 
The SDE is, by definition, a poor posed statistic to characterize the
reliability of a signal.
The standard deviation of the test statistic is included in the
denominator of the SDE. 
Therefore, a very significant peak, such as that created by a transiting
hot Jupiter, will have a significant contribution to the standard
deviation of the data, therefore reducing the significance of the SDE
\citep[see the discussion in][]{kovacs2002}.
If we compare just the value of the test statistic of DST, it has a
value of 3.5 in the peak and a background of 0.004 in the
case of CoRoT-1b (a ratio of 800) and a value of 0.023 in the
peak and a background of 0.0006 in the case of CoRoT-7b (a ratio of 38).
As expected, the DST statistic reflects correctly that CoRoT-1b is a
more significant detection than CoRoT-7b.
This is again a consequence of the definition by \citet{kovacs2002}
of the SDE statistic. 

%
\section{The transit search paradigm}
\label{sec:paradigma}

\subsection{The period array}
\label{subsec:period}

The most common strategy to search for transit-like signals is to
perform a blind analysis for an array of test periods. 
The extremes of the period array are defined according to the expected
detectability range: periods shorter than 0.5 days are not expected
for planets around solar-like stars because of the vicinity to the
host star (as discussed in the introduction).
On the other hand, the longest detectable period depends on the length 
of the observations and the duty cycle. 
Finally, one has to define the density of periods in the array. 
Not all authors describe their choices in the literature.
Two exceptions are \citet{clarkson2007}, who for the
SuperWASP-North survey defined a period range between 0.9 and 
5 days with a spacing of 0.002 days for a coarse search and of 0.001
days for a finer search, and \citet{faedi2011}, who used a method
equivalent to ours, although in the frequency range.
See also the discussion in \citet{jenkins1996,jenkins2010c}.

The light curve is then folded to each value of the period. 
For each of these periods, the detection algorithm is run for every
reasonable combination of transit epoch (from 0 to the end of the
folded light curve) and duration of the transit (which can be
estimated, for example, from the expected relative sizes of planets
around solar-like stars using the expressions in
\citealt{seager2003}). 
The spacing in the arrays of epochs and durations can be adjusted
considering the number of bins in the fold. 
This method implicitly assumes that there is only one transit per
folded light curve.

We propose a method to estimate the best density of the search
arrays and an alternative to the blind search described above.

The region in transit $T$ of a folded light curve is defined as the
points in the interval $[x_0-d/2,x_0+d/2],$ where $x_0$ is the epoch
of the transit and $d$ its duration. 
If the light curve is not folded, the equivalent definition of the
region in transit is the set of points that verify the condition:
\begin{equation}
\label{eq:region_in_transit}
x - \mathrm{floor}\left( x/P\right) \cdot P \in [x_0-d/2,x_0+d/2].
\end{equation}

If we determine the period $P$ of a transiting planet with an error
$\Delta P$ and its epoch $x_0$ with an error $\Delta x_0$, we can
constrain the position of the $N$th transit $x_N$ with an error
$\Delta x_N = \Delta x_0 + N \cdot \Delta P$. 
The position of the $N$th transit depends on the length of the
observations $I$ and typically $N \sim I/P.$
Therefore, if we aim to constrain the position of the $N$th transit
with an uncertainty smaller than a certain value $h$ (for example,
comparable to the sampling rate), the acceptable uncertainty in the
period has to fulfill the condition 
$\Delta P < h*P/I.$
This condition puts a physical constraint on the desired density of 
the period array.

\subsection{The region of interest}
\label{subsec:roi}

The region of interest obviously is the transit.
The most simple strategy is to search one transit-like feature per
period, neglecting the presence of an occultation or secondary
transit, which is indeed a reasonable assumption for ground-based
surveys, and neglecting the presence of transiting systems of
planets.
The \dst\, model allows (see definition \ref{eq:function_Dst})
to search for $k$ transit-like features simultaneously without the
need of folding the light curve.
The definition of the region in transit $T_k$ for the $k$th transit
will be equivalent to (\ref{eq:region_in_transit}), substituting $x_0$ 
by $x_0^k$, because each transit-like feature will have a different
epoch.
One can also use this system to search for secondary
transits\footnote{the computational effort can be reduced by assuming
  circular orbits and fixing $x_0^2=x_0^1+P/2,$ but this assumption is
  only reasonable for short periods.} by setting $k=2$.
Including a second transit-like feature in the model is not a
significant improvement for planet surveys, because planetary
occultations are orders of magnitude lower than primary occultations
(the temperature contrast between the star and the planet is big) and
therefore the improvement of the test statistic will be small.
But it can be justified in the case of eclipsing binaries, where the
temperature contrast is comparable and therefore primary and secondary 
eclipse have comparable depths.
This approach improves the detection and characterization of eclipsing
binaries, which are the main source of contamination in the search for
transiting planets.

\subsection{Transit timing variations}
\label{subsec:ttv}

\begin{figure}[t]
  \begin{center}
    \includegraphics[%
      width=0.9\linewidth,%
      height=0.5\textheight,%
      keepaspectratio]{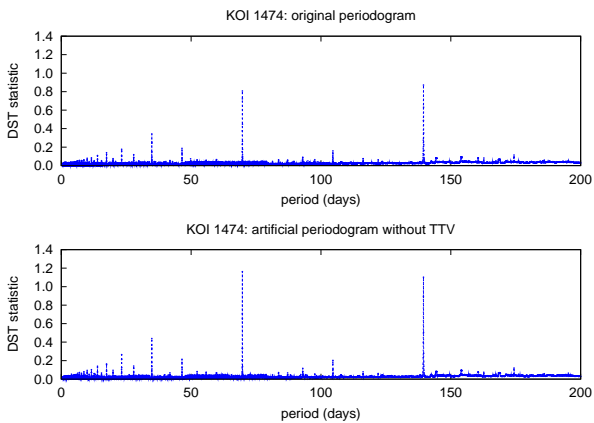}
  \end{center}
  \caption{
    Comparison of the DST detection statistic for KOI~1474 in the
    original data set (top) and once the transit timing variations
    were artificially removed (bottom).
  }
  \label{fig:KOI1474}
\end{figure}

\begin{figure}[t]
  \begin{center}
    \includegraphics[%
      width=0.9\linewidth,%
      height=0.5\textheight,%
      keepaspectratio]{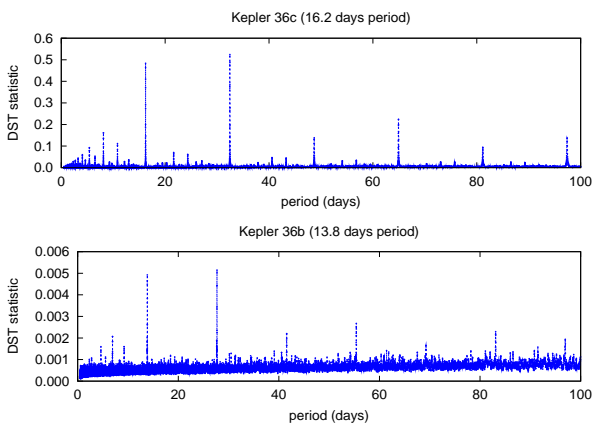}
  \end{center}
  \caption{
    Comparison of the DST detection statistic for the planets of the
    Kepler-36 system. Kepler-36c (above) with a size of 3.7 Earth
    radii and a period of 16.2 days produces a clear detection signal
    in the periodogram. Kepler-36b (below) with a size of 1.5 Earth
    radii and a period of 13.8 days, which shows significant transit
    timing variations, produces a distinguishible signal, although
    limited by the non-periodicity of the transit.
  }
  \label{fig:kepler36}
\end{figure}

The shape of a transit is markedly affected when folding the
light curve of a planet with significant transit timing variation (TTV)
assuming a constant period (see, for example, the case of Kepler-9c,
\citealt{holman2010}). 
Although our algorithm performs reasonably well when detecting planets
or planet candidates with significant TTVs, the current configuration
has a limited performance. 
See, for example, the case of KOI~1474 \citep{dawson2012}, a
transiting Jupiter-sized candidate with a period of $\sim 70$ days
and TTVs of up to 1h from transit to transit.
We analyzed the original \kepler\,Q0 to Q6 data and calculated
the periodogram (see upper part of Fig~\ref{fig:KOI1474}) and the
TTVs. 
We used the value of the TTVs to artificially shift each transit from 
the original data to its expected position, if the orbit was exactly
periodic, and we recalculated the DST statistic (bottom part of
Fig~\ref{fig:KOI1474}).
It now shows a value at the period of the candidate 25\% higher.
To overcome this effect, we propose another possibility that 
optimizes the search for planets with significant timing variations.
If the observations span an interval $I$ and we are searching
for a planet with a period $P$, we expect $k=I/P$ transits (assuming
a full duty cycle). 
We can define $k$ regions of search and minimize for each region with
respect to the epoch of the $k$th transit, which would be treated
independently in the case of large timing variations.
Each region $k$ will cover a region of length $P$ since the end of the
previous region $k-1$.
This increases the computational effort, 
but it provides the best description for this case of scenarios. 
In the case of Kepler-9c the S/N ratio of the detection
was high enough to surmount the degradation of the transit signal.
But this might not be the case for small planets close to the noise
level. 
For a given perturber, the timing variations will be stronger for
smaller planets \citep{agol2005,holman2005,borkovits2011}. 
Recently, \citet{carter2012} mentioned a transit detection algorithm
that accounts for variations between consecutive transits, similar
to the idea described above, although the algorithm is not described
in detail in their paper.
We analyzed \kepler\,public data from quarters Q0 and Q9 with DST
and are able to detect both planets with reasonable S/N ratio,
although the detection of Kepler-36b is indeed limited by the
presence of the TTV (see Fig.~\ref{fig:kepler36}).

\subsection{False alarms}
\label{subsec:fa}

Every transit detection algorithm based on a test statistic to
determine the presence of a transit must use a threshold
to distinguish between real and spurious signals.
If the probability distribution function of the test statistic is
known, it is straightforward to fix the threshold by allowing any
desired value of spurious false positives.
This procedure is widely discussed in the literature in different contexts 
\citep[see for example][]{schwarzenberg-czerny1989,jenkins2002a}.
However, in practice these limits are not easily applicable.
The fundamental assumption that a given statistic
follows a particular probability distribution function is that the
errors affecting the observations can be treated as random 
Gaussian variables. 
Unfortunately, as we show below, this is not the case because
most of the remaining residuals, and for sure the most annoying, are
always correlated to some degree, invalidating the previous
assumption.
Therefore, the threshold preventing the appearance of false alarms has
to be set accordingly to the level of correlated noise remaining in
the light curves, which is method dependent, and not merely to the
expected value according to the theoretical behavior of the
statistic.
All false alarms found in this paper are related to residuals of
the stellar activity or instrumental residuals (see below).

%
\section{Future developments}
\label{sec:futuro}

The previous section showed the improvements in the planetary
detection performance brought by the better description of the transit
shape and the better test statistic for both giant and terrestrial
planets.
The only drawback of the \dst\, formalism in comparison to BLS is that
it involves a slightly heavier computational effort.
However, this is only a technical question, not a formal one, and
moreover the increase in computational time is not prohibitive. 
Another advantage of our model is that it keeps the same number of 
free parameters as BLS, not like the trapezoidal model.
The drawback of the computational effort is only an inconvenience and 
perhaps can be overcome by changing the search strategy.
The blind search requires to try, for every trial period, every epoch
and duration (or every $k$th epoch and duration), which is quite
expensive and becomes worse with longer exposure intervals, such as
those of the \kepler\,and PLATO surveys.
We can overcome this difficulty by changing the paradigm of the
search. 
One can assume that each point of the light curve is the center of a 
transit-like feature of a given length, therefore defining as many
transits as points in the light curve ($k=N$).
Then one can calculate the value of the expressions in
(\ref{eq:Dst_Lambdas}) for each point and choose which subset of
points of the light curve produces a significant change of the test
statistic (\ref{eq:Dst_estimador}).
If this subset of points corresponds to a periodic signal, or even
multiperiodic in the case of several transiting planets in the system,
we have a positive detection. 
However, one cannot yet find such algorithms in the literature that 
can select the subset of points that produces a significant change of
the test statistic. 
Searching for these particular algorithms is beyond the scope of this 
paper, which only aims to present the advantages of the model
described in (\ref{eq:function_Dst}), but it will be the subject of
future study.

%
\section{Data treatment}
\label{sec:datos}

We analyzed Q1 data from the \kepler\,mission. 
In particular, we use the raw flux column ({\em ap\_raw\_flux})
described in the \kepler\,data documentation
\footnote{In the latest released version of \kepler\,data, the raw flux
  column is called {\em sap\_flux}. See {\tt
    http://keplergo.arc.nasa.gov/Documentation.shtml}}. 
We did not use any filter for outliers, relying on the \kepler\,pipeline 
for the filtering of cosmic rays.
We analyzed each light curve with a Lomb-Scargle periodogram
\citep{scargle1982} to remove harmonic stellar pulsations.
We then applied the stellar variability filter detailed below before
using the transit detection algorithm DST to search for the planetary
candidates. 

\subsection{Stellar variability filter}
\label{subsec:stellarVariabilityFilter}

The complex variability pattern of solar-like stars has been studied
in the framework of space-based transit surveys such as \corot\, and 
\kepler\,\citep[see, for example][]{aigrain2009,gilliland2011}.
The stellar variability filter described here is optimized for the
removal of the relatively slow changes caused by the evolution of
stellar spots modulated by the rotational period of the star. 
The amplitudes involved in this kind of activity are of a few
percent in flux with time-scales of several days.
This pattern is regularly observed by space-based telescopes, where
targets such as HD~189733 observed by MOST \citep{croll2007}, CoRoT-2 
\citep{alonso2008a}, or HAT-P-11 observed by \kepler\,
\citep{deming2011,sanchisojeda2011} are characteristic examples.
Other activity patterns such as granulation and related phenomena,
which have typical amplitudes in of 0.01\% to the ppm range and
frequencies of hundreds of $\mu$Hz, are not treated here.


We assumed that we can separate the incoming flux from the target into
a stellar signal and a planetary transit signal plus some
non-correlated residual noise:
\begin{displaymath}
  F(t) = F_\mathrm{star}(t) + F_\mathrm{transit}(t) + w(t).
\end{displaymath}
This neglects any contribution from instrumental features, which in
reality will produce correlated residuals. 
In practice, the data were pre-treated to remove any previously known 
instrumental signal,\footnote{for example, in the \corot\,context, any
  periodical variation at the orbital period of the satellite.}
and any remnant feature will remain in the residuals (which therefore
would not be treatable as random Gaussian variables, as
discussed above).
We assumed that the stellar signal $F_\mathrm{star}$ presents some
spot-induced activity signature, which has typical time-scales of days
and amplitudes of few percent; the planetary transit contribution 
$F_\mathrm{transit}$ is zero everywhere except during the transits,
where it has the characteristic shape of a transiting planet
\citep[see, for example][]{mandel2002}.
The time-scale of a planetary transit, which is of some hours, 
is different from the typical time-scales of the spot induced
stellar variability, therefore we aimed to define an algorithm that can
separate both.
To build a model of the light curve that is sensitive only
to the long-term stellar variability, we binned the data adding up to 
$n_\mathrm{bin}$ points per bin.
This number was chosen for the purpose of averaging out as much as
possible the signatures of transits while the number of points in the 
model was dense enough to resolve the stellar variability.

Subsequently we used the Savitzky-Golay approach, which consists in
assigning a polynomial model to the original data points that fits the
binned data better in a least-squares sense 
(see \citealt{savitzky1964} and \citealt{press2002}).
We modeled up to $n_\mathrm{scale}$ binned data points with a set of 
Legendre polynomials of degree $n_\mathrm{Leg}$.
The compromise found between the number of points in the bins
$n_\mathrm{bin}$, the number of points modeled each iteration
$n_\mathrm{scale}$ and the degree of the polynomials $n_\mathrm{Leg}$
governs the reaction of the model to the stellar variability and the
presence of transits.

\begin{figure}[t]
  \begin{center}
    \includegraphics[%
      width=0.9\linewidth,%
      height=0.5\textheight,%
      keepaspectratio]{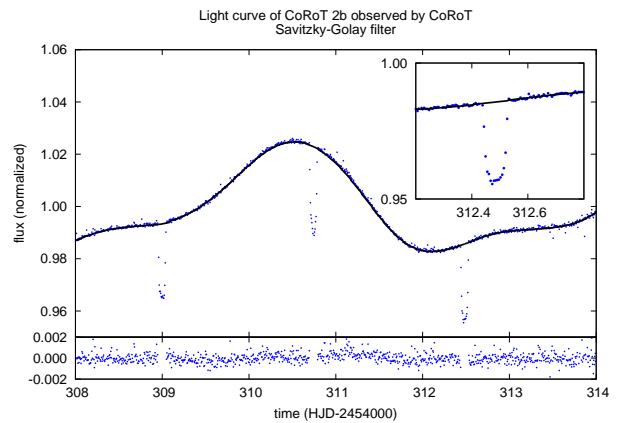}
  \end{center}
  \caption{
    Model of the stellar activity of the \corot\, light curve of
    CoRoT-2b. The inner box shows the detail of the modeling around
    one of the transits. The lower box shows the residuals of the
    modeling. 
  }
  \label{fig:c2bmtp}
\end{figure}

\begin{figure}[t]
  \begin{center}
    \includegraphics[%
      width=0.9\linewidth,%
      height=0.5\textheight,%
      keepaspectratio]{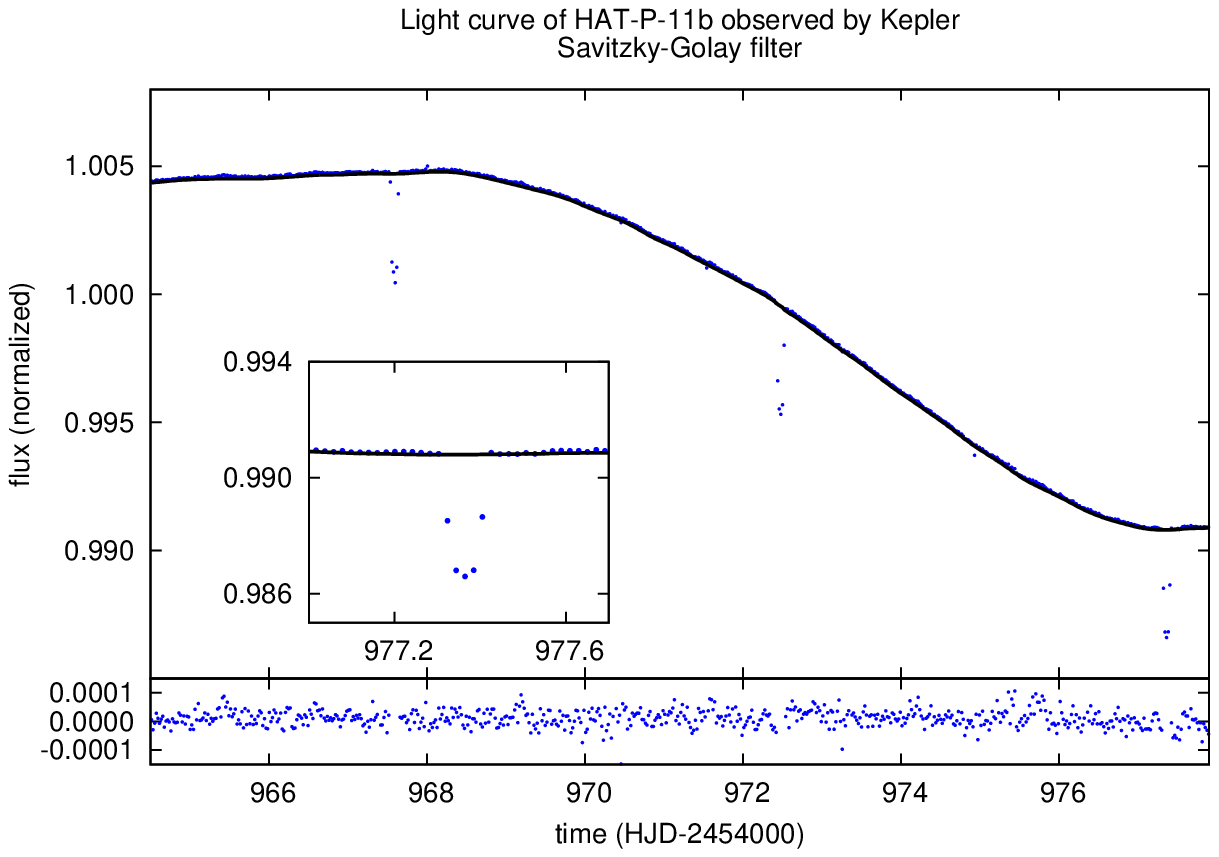}
  \end{center}
  \caption{
    Model of the stellar activity of the \kepler\,light curve of
    HAT-P-11b. The inner box shows the detail of the modeling around
    one of the transits. The lower box shows the residuals of the
    modeling. 
  }
  \label{fig:h11bmtp}
\end{figure}

A practical procedure to minimize the effect of transits and outliers
is to remove the $n_\mathrm{rem}$ worst points of the fit of the
polynomial and recalculate the coefficients of the polynomial with
$n_\mathrm{scale}-n_\mathrm{rem}$ points.
This procedure can be repeated several times ($n_\mathrm{times}$).
Finally, we built $F_\mathrm{star}$ with the values of the polynomials
at every time measurement.
The values for the different parameters used in this work are
summarized in Table~\ref{table:sgparam} and the performance of the
filter is shown in two examples named above: CoRoT-2b and HAT-P-11b 
(Figs.~\ref{fig:c2bmtp} and \ref{fig:h11bmtp}). 

\begin{table}
  \caption{\label{table:sgparam}Configuration parameters of the Savitzky-Golay filter.} 
  \renewcommand{\arraystretch}{1.2}
  \begin{tabular}{lcc}
    \hline\hline
    parameter & Kepler data & \corot\, data \\
    \hline
    $n_\mathrm{bin}$   &  5 & 20 \\
    $n_\mathrm{scale}$ & 11 & 11 \\
    $n_\mathrm{Leg}$   &  5 &  5 \\
    $n_\mathrm{rem}$   &  1 &  1 \\
    $n_\mathrm{times}$ &  3 &  3 \\
    \hline
  \end{tabular}
\end{table}

The parameters of the filtering method used were optimized for \corot\,
data and may not be ideal for the analysis of \kepler\,data. 

\begin{figure}[t]
  \begin{center}
    \includegraphics[%
      width=0.9\linewidth,%
      height=0.5\textheight,%
      keepaspectratio]{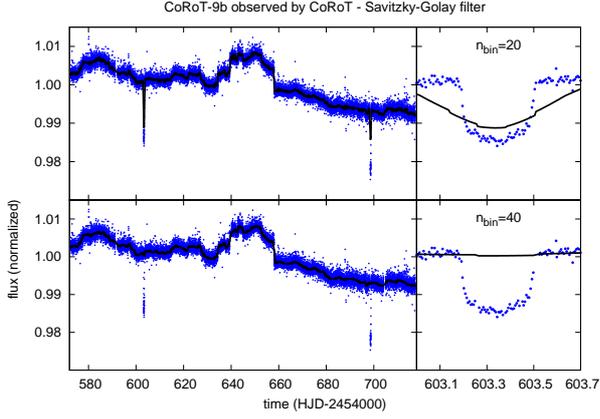}
  \end{center}
  \caption{
    Model of the stellar activity of the light curve of
    CoRoT-9b with two different values for the configuration of the
    stellar activity filter. In the top panel, the configuration of
    the filter is not optimal and the filter tries to remove the
    transit signal. In the lower panel, the configuration was
    adapted to preserve the signal of these long transits.
  }
  \label{fig:corot9b}
\end{figure}

However, those parameters were optimized to preserve
the shape of transits of short-period planets, such as the ones we
expect to detect in the \kepler\,Q1 data set. 
They are not optimal for the search of planets with longer periods
because they distort the shape of long transit features.
For example, in the case of CoRoT-9b (shown in Fig.~\ref{fig:corot9b})
we needed to use 40 points per bin to preserve the shape of the
transit. 

Note that the nominal sampling rate of \corot\, is 512s instead of the
nominal sampling rate of 30 minutes of \kepler, therefore
$n_\mathrm{bin}=20$ in the \corot\, sampling rate corresponds
approximately to $n_\mathrm{bin}=5$ in the \kepler\,sampling rate.

\begin{figure}[t]
  \begin{center}
    \includegraphics[%
      width=0.9\linewidth,%
      height=0.5\textheight,%
      keepaspectratio]{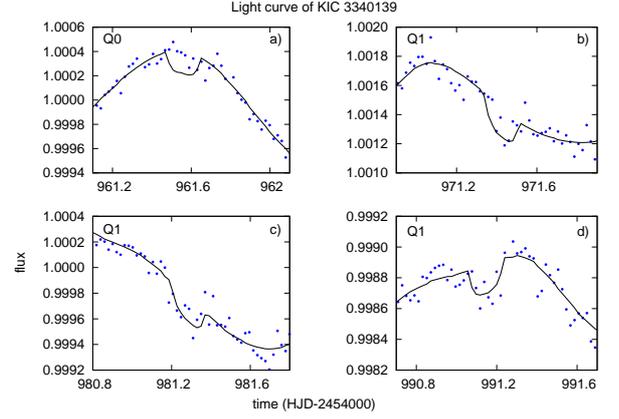}
  \end{center}
  \caption{
    Detail of the light curve of KIC~3340139 observed in the Q0 and Q1
    \kepler\,quarters. 
    The continuous line represents the model of stellar activity
    together with the transit solution found by DST.
    Panels b-d show residuals of stellar activity that mimic
    a periodic transit feature. 
    Panel a shows the expected position of the transit in Q0
    data.} 
  \label{fig:kic3340139_mtp}
\end{figure}

\begin{figure}[t]
  \begin{center}
    \includegraphics[%
      width=0.9\linewidth,%
      height=0.5\textheight,%
      keepaspectratio]{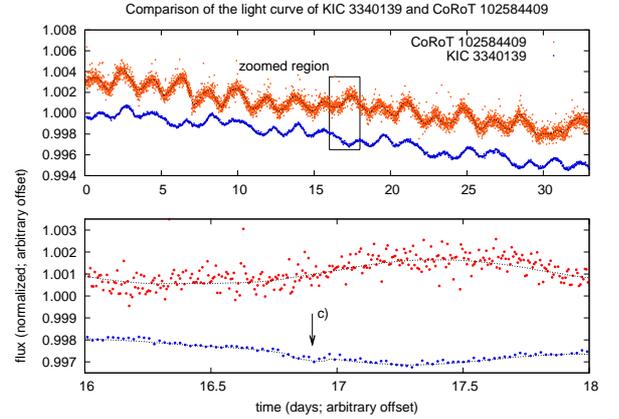}
  \end{center}
  \caption{
    Comparison of the light curves of KIC~3340139 and CoRoT 102584409,
    which show similar activity levels. The position of event c from 
    Fig.~\ref{fig:kic3340139_mtp} is marked with an arrow in the lower
    panel.}
  \label{fig:kic3340139_CoRoT102584409_mtp}
\end{figure}

\begin{figure}[t]
  \begin{center}
    \includegraphics[%
      width=0.9\linewidth,%
      height=0.5\textheight,%
      keepaspectratio]{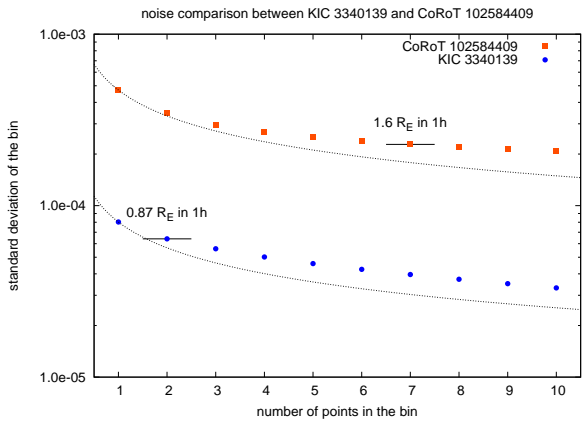}
  \end{center}
  \caption{
    Analysis of the noise level of the light curves of KIC~3340139 and
    CoRoT 102584409. The typical noise level within 1h in the \corot\,
    light curve is equivalent to the depth of a transit of a $1.6R_E$
    planet around a solar-like star. The typical noise level for the
    \kepler\,light curve in 1h is equivalent to the depth of a transit
    of a $0.87R_E$ planet. The dotted line shows the $1/\sqrt{x}$
    expected white noise trend.}
  \label{fig:kic3340139_CoRoT102584409_rednoise}
\end{figure}

On the other hand, \kepler\,reaches a higher S/N ratio level than
\corot\, and many stellar activity patterns, which are too small to be
important for  \corot\,, become critical for \kepler. 
This is shown in Fig.~\ref{fig:kic3340139_mtp}. 
The Q1 light curve of KIC~3340139 shows three events approximately
separated by 9.86 days (labeled with the letters b, c, and d in the
figure). 
If they were produced by a transiting planet around a solar-like star,
the size of the planet could be as small as 1.5 Earth radii.
However, those events are not convincing enough: 
event b is in fact triggered by only three points $\sim 2\sigma$
below the mean.
Event c is probably caused by a residual of the stellar activity
on a shorter timescale than the spot-induced variability.
Event d is more similar to a planetary transit event, but the
ingress and egress are not well defined and indeed there are similar
features in the light curve which do not follow any particular
periodic pattern.
Event a corresponds to Q0 data and clearly shows no 
transit-like feature at the expected position of the transit.
We believe that the periodical signal detected by the algorithm in
KIC~3340139 is an artifact caused by residuals of the stellar activity
that wer not properly handled by the filter because of their
short typical timescales, comparable to the duration of a planetary
transit.
An important indication against the planetary origin of the events is
that if the configuration of the stellar activity filter is changed,
for example making $n_\mathrm{bin}=8$, no periodical signal is
detected in the light curve any more. 
Any detection that depends so strongly on the filtering method is not
reliable.
From another standpoint, the activity level displayed by KIC~3340139
is not unusual.
The light curve of the LRa01 run from CoRoT 102584409 (a star with a
similar magnitude to KIC~3340139, r'=11.44 mag and K$_p$=11.46 mag
respectively) shows a similar activity pattern 
(see Fig.~\ref{fig:kic3340139_CoRoT102584409_mtp}), but \corot\, is
not sensitive to the small amplitude, short timescale patterns such as
event c in Fig.~\ref{fig:kic3340139_mtp}.
Figure~\ref{fig:kic3340139_CoRoT102584409_rednoise} compares the noise
level of these particular \corot\, and \kepler\,targets following the
procedure described by \citet{pont2006}, which separates the
contributions of the non-correlated (also known as white) noise,
which evolves as $1/\sqrt{n}$, being $n$ the number of binned flux
values, and the correlated (also known as red) noise.
The filter applied here was optimized for \corot\, data and is
insensitive to the patterns displayed by KIC~3340139. 
We consider optimizing the algorithm filtering the stellar activity
for small \kepler\,transits in future work.

This limitation of the stellar activity filtering created a
considerable number of false alarms in this data set compared to what
is usual when analyzing \corot\, data (more than 100), which were removed
by running the filtering method again with slightly different
parameters and by comparing the results of the transit detection
analysis in the new filtered data.
Any detection that depended on the filtering method was rejected
as a false alarm. 
As these are dependent on the stellar filtering method applied and not
on the detection method, they are not listed in this paper.

%
\section{Results}
\label{sec:resultados}

As discussed before, one of the main conclusion of the studies carried
out by the \corot\,community when testing different transit detection
algorithms was that both detections and false alarms were method
dependent \citep{moutou2005,moutou2007}.
In our analysis of the Q1 data set we found that both statements
are still true when comparing our results to those from the \kepler\,
community.

\subsection{The yield of detections}
\label{subsec:yieldOfDetections}

\begin{figure}[t]
  \begin{center}
    \includegraphics[%
      width=0.9\linewidth,%
      height=0.5\textheight,%
      keepaspectratio]{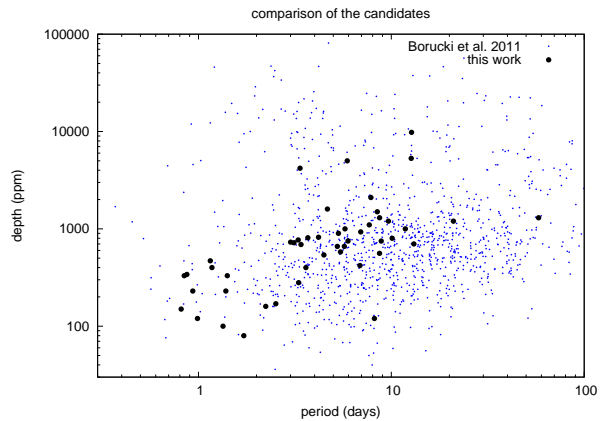}
  \end{center}
  \caption{
    Comparison of the depth and the period of the planetary candidates
    found in this work compared to those of \citetalias{borucki2011b}.
  }
  \label{fig:depth_vs_per}
\end{figure}

We found \noNewCandidates\,planetary candidates analyzing 33 days of
data that went unnoticed to \citetalias{borucki2011b}, among 
them \noNewSystems\,multiple transiting systems.
Most of them were discovered later by \citet{batalha2012}, who
analyzed 480 days of data, but not all of them.
There are still \noNewCandidatesBatalha\, planetary candidates, among
them \noNewSystemsBatalha\,multiple transiting systems, not discovered
and published by the \kepler\, team\footnote{while 
  our paper was in the revision process, \citet{ofir2012} published 84
  new transiting signals in the \kepler\, data.}. 
The advantage of having an algorithm that can detect new candidates in
a significantly reduced data set is two-fold. 
On the one hand, the fewer transits needed to find a planet,
the earlier can the follow-up begin, which is a critical point in the 
design of transit surveys because the resources are limited.
On the other hand, the most interesting planets, such as terrestrial
planets in the habitable zone of solar-like stars, have long periods,
typically comparable to years, and will typically have few transit
events.
If we compare the distribution of our new candidates with those from 
\citetalias{borucki2011b}, we can see that the bulk of our detections
has periods between five and ten days (see
Fig.~\ref{fig:depth_vs_per}), which corresponds to candidates having
six to three transit events in our data set.

\addtocounter{table}{1}

Table~\ref{table:candidates} contains the ephemeris of the new
planetary candidates not found by \citetalias{borucki2011b}, nor found
as eclipsing binaries by \citet{prsa2011} or \citet{slawson2011}.
Some of them were later found by \citet{batalha2012}; when this is the
case, the KOI number assigned by the \kepler\, team is shown to identify
them. 
These candidates were ranked according to the \corot\, rules (see
\citealt{carpano2009,cabrera2009}).
Note the presence of several planetary candidates similar to
CoRoT-7b\,(\citetalias{leger2009}; \citealt{queloz2009}) and 
Kepler-10b\,\citep{batalha2011} in period and depth (they are marked
with the label C7b-like). 
To detect the candidates we only used data from Q1, but to
characterize the candidates and calculate the ephemeris we used public
\kepler\, data from other runs to improve the ephemeris and confirm
the detection.

\addtocounter{table}{1}

Table~\ref{table:fp} contains the ephemeris of our false positives:
detections ranked by the \corot\,rules that turned out to be
background eclipsing binaries identified by the study of the centroid
motion \citep{batalha2010b}. 
The responsible contaminating eclipsing binary or its position within
the mask is indicated in the table.

\addtocounter{table}{1}

Finally, table~\ref{table:fa} contains the ephemeris of the false
alarms found in our analysis of the Q1 data, which are discussed below
in subsection \ref{subsec:falseAlarms}.

\subsection{Kepler candidates not detected by our pipeline}
\label{subsec:nonDetections}

We studiied the candidates found in \citetalias{borucki2011b} that 
were not found by our analysis and discuss in the following
sections possible reasons for these non-detections.
Both data sets are not immediately comparable, because
\citetalias{borucki2011b} corresponds to four times more data than the
analyzed here, and therefore we do not expect to find all. 
Still, we can learn about the performance of our algorithm and about
the \kepler\,data set from these non-detections, as we detail below. 

\subsubsection{Candidates below the detection threshold}
\label{subsubsec:nonDetectionsBelowThreshold}

\begin{figure}[t]
  \begin{center}
    \includegraphics[%
      width=0.9\linewidth,%
      height=0.5\textheight,%
      keepaspectratio]{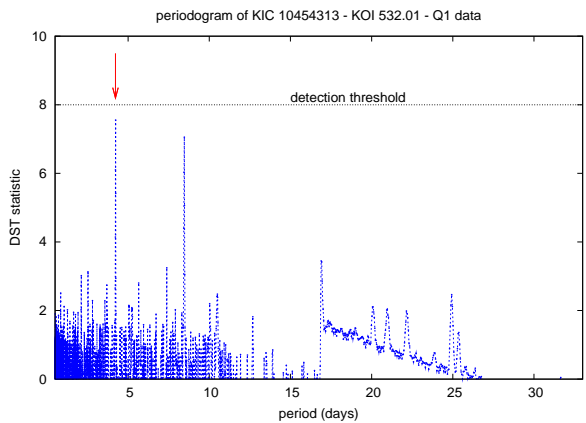}
  \end{center}
  \caption{
    Periodogram of KIC~10454313 showing the position of the peak
    corresponding to the planet candidate KOI~532.01, marked with an
    arrow. The height of the peak is 7.9, directly below the detection
    threshold (see discussion in text).}
  \label{fig:kic10454313_pdg}
\end{figure}

\begin{figure}[t]
  \begin{center}
    \includegraphics[%
      width=0.9\linewidth,%
      height=0.5\textheight,%
      keepaspectratio]{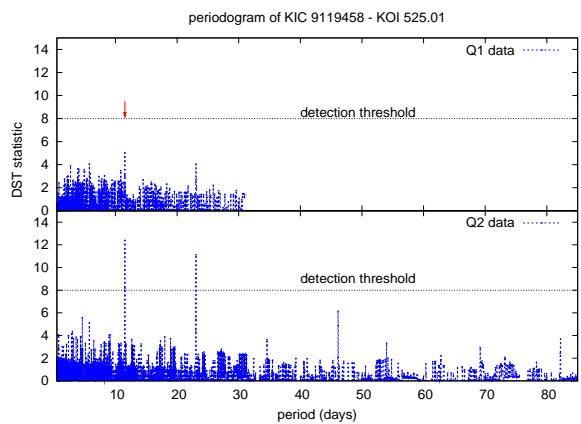}
  \end{center}
  \caption{
    Periodogram of KIC~9119458. The upper part corresponds to the
    analysis of the data set Q1 and the peak of the candidate 525.01
    is marked with an arrow. Its height is 5.4, well below the
    detection threshold. The lower part corresponds to the analysis of
    the data set Q2, where the height of the peak is 12.8 (see
    discussion in text).}
  \label{fig:kic9119458_pdg_mtp}
\end{figure}

In several cases, we found \citetalias{borucki2011b} \kepler\,candidates
below our detection threshold.
The threshold was optimized according to the \corot\,experience and
was fixed at a S/N ratio of 8 between the peak and the background.
Several candidates, which could be clearly detected in the
light-curves, produced periodograms with peaks immediately below the
fixed threshold (see, for example, the case of the candidate
KOI~532.01 from \citetalias{borucki2011b} shown in
Fig.~\ref{fig:kic10454313_pdg}). 
In the absence of red noise, this detection threshold can be fixed for
a particular detection algorithm by making the probability of
detecting a false alarm as low as needed 
\citep[see, for example,][]{jenkins2002b}. 
However, in practice the presence of correlated residuals in the light
curves distorts the expected distribution of false alarms and it is
advisable to use a more conservative value of the threshold.
An illustrating case is shown in Fig.~\ref{fig:kic9119458_pdg_mtp},
where the Q1 data set alone, which includes three transit events,
barely suffices to distinguish between the genuine peak from the
candidate KOI~525.01 from \citetalias{borucki2011b}. 
For comparison, in the lower part of Fig.~\ref{fig:kic9119458_pdg_mtp}
the analysis of the data set Q2, which contains seven transit events,
is shown. 
In this data set the peak clearly rises above the fixed threshold.
Considering the level of the correlated residuals in the Q1 data set,
the threshold applied in the analysis presented in this paper seems to
be too conservative.

\subsubsection{Instrumental effects}
\label{subsubsec:nonDetectionsInstrumentalEffects}

\begin{figure}[t]
  \begin{center}
    \includegraphics[%
      width=0.9\linewidth,%
      height=0.5\textheight,%
      keepaspectratio]{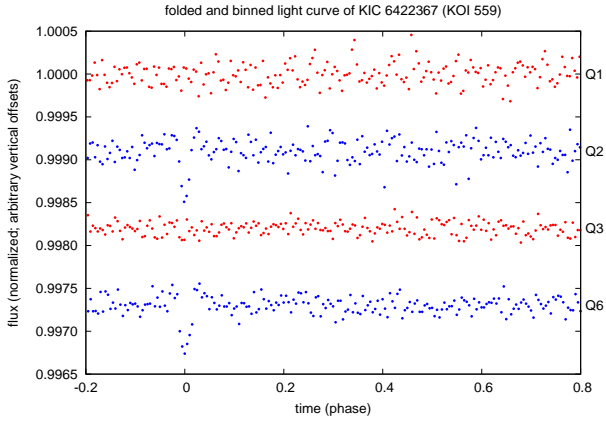}
  \end{center}
  \caption{
    Comparison of the light curves of the target KIC~6422367 folded at
    the period and phase of the planetary candidate KOI~559.01. The
    signal of the candidate is not visible in quarters Q1, Q3, and
    Q4 (not shown). See discussion in text. }
  \label{fig:kic6422367_transit_Q1Q2Q3Q6}
\end{figure}

\begin{figure}[t]
  \begin{center}
    \includegraphics[%
      width=0.9\linewidth,%
      height=0.5\textheight,%
      keepaspectratio]{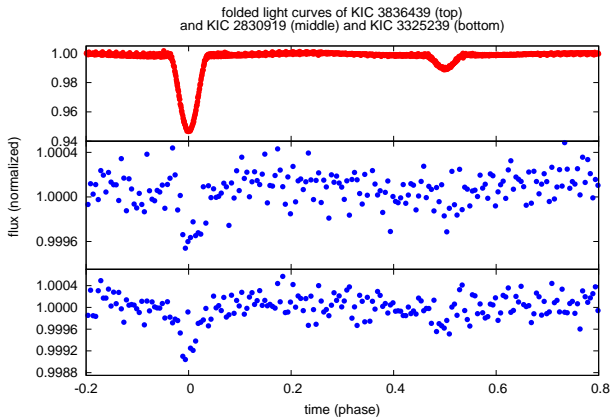}
  \end{center}
  \caption{
    Comparison of the light curves of the targets
    KIC~2830919 and KIC~3325239 folded at the ephemeris of the
    eclipsing binary KIC~3836439. See discussion in text. }
  \label{fig:kic2830919_kic3325239_vs_kic3836439}
\end{figure}

\begin{figure}[t]
  \begin{center}
    \includegraphics[%
      width=0.9\linewidth,%
      height=0.5\textheight,%
      keepaspectratio]{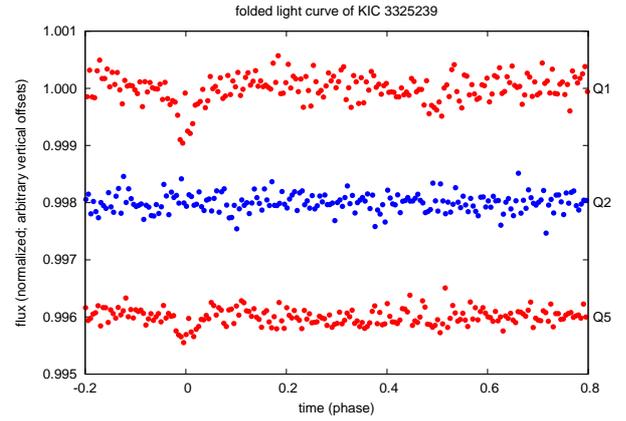}
  \end{center}
  \caption{
    Comparison of the light curves of the target KIC~3325239 folded at
    the period and phase of the eclipsing binary KIC~3836439. From top
    to bottom the data of quarters Q1, Q2, and Q5 are shown. The
    signal is only visible in quarters Q1 and Q5. See discussion in
    text. }
  \label{fig:kic3325239_transit_Q1Q2Q5}
\end{figure}

\begin{figure}[t]
  \begin{center}
    \includegraphics[%
      width=0.9\linewidth,%
      height=0.5\textheight,%
      keepaspectratio]{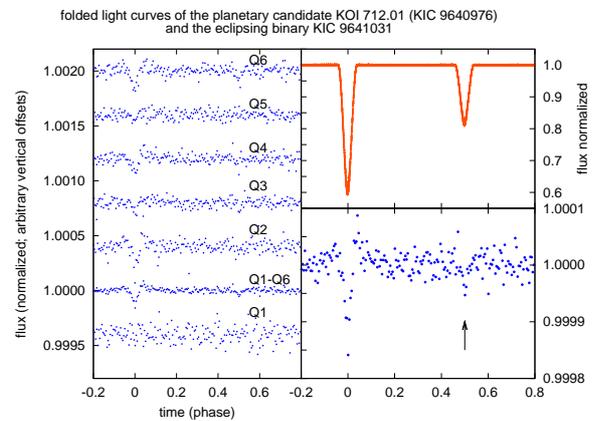}
  \end{center}
  \caption{
    Comparison of the light curves of the planetary candidate
    KOI~712.01 (KIC~9640976) and the eclipsing binary KIC~9641031. The 
    left side of the figure shows the data of KOI~712.01 from the
    different quarters. The transit events are not seen in Q1, and the
    transit depth is not constant in quarters Q2 to Q6. The top part
    of the right side shows the light curve of the eclipsing binary
    and the lower part the data from quarters Q1 to Q6 of 712.01
    folded at the ephemeris of the eclipsing binary. The secondary
    eclipse is visible in this data set (marked with an arrow in the
    figure). }
  \label{fig:koi712_kic9641031}
\end{figure}

Among the candidates not found in our analysis there is a very
particular case that is worth mentioning.
The planet candidate from \citetalias{borucki2011b} KOI~559.01
(KIC~6422367) was not found in the Q1 data set. 
We looked for the candidate in the Q2 data set and we were able to
find it, but with a significantly different depth, $700\pm70$ ppm,
than the one  reported in the original paper ($214$ ppm).
We pursued our analysis on this particular candidate by analyzing the
latest data sets released by the \kepler\,team (see 
Fig.~\ref{fig:kic6422367_transit_Q1Q2Q3Q6}) and realized that the
candidate was only present in the Q2 and Q6 data sets, not in Q1,
Q3 and Q4 (there is no Q5 data for this target because of the failure
of module 3). 
We note that from Q2 to Q6 the satellite performs a full revolution
maneuver. 
This behavior is typical of a contamination from a background
eclipsing binary, which only pollutes the PSF of the candidate in  
particular orientations of the CCD.
We searched for an eclipsing binary that could be the origin of this
contamination in the field of view of \kepler\,and found that
KIC~6422367 has the same period and roughly the same phase as the
candidate. 
The periods of the candidate and the binary are 
$4.331\,39\,(5)$ days and $4.331\,398 (1)$ days respectively; the
epochs are (HJD-$2\,451\,545.0$) $121.722\,(4)$ and
$121.732\,21\,(7)$. 
The main counterargument of this explanation is that this eclipsing
binary is $1.8^{\circ}$ away from the candidate (for example, during
Q2, the eclipsing binary was observed in module 11 in channel 34,
whereas the planetary candidate was observed in the same module, but
in channel 36).
The probability that two independent targets have the same period and
phase is extremely low.
On the other hand, this phenomenon could have an instrumental origin. 
In support of this hypothesis, we have found two more examples of this
behavior.
The targets KIC~2830919 and KIC~3325239 show the same ephemeris as the
eclipsing binary KIC~3836439 
(see Fig.~\ref{fig:kic2830919_kic3325239_vs_kic3836439}).
The first target is at $1.4^\circ$ of the binary and the second at
$0.8^\circ$.
KIC~3325239 clearly shows the alternates of appearance in quarters
Q1 and Q5, while it disappears in quarters Q2, Q3, Q4 and Q6 
(see Fig.~\ref{fig:kic3325239_transit_Q1Q2Q5}). 
In space-based surveys, the contamination of neighbor targets by a
bright eclipsing binary, such as the contamination of KIC~9641031 of
the targets KIC~9640985, KIC~9641008, KIC~9641041, and the planet
candidate KOI~712.01 (KIC~9640976, which is then a false alarm despite
of being ranked as a planet candidate in \citetalias{borucki2011b},
see Fig.~\ref{fig:koi712_kic9641031}), is well known (see, for
example, \citealt{deeg2009,batalha2010b}), but to our knowledge the 
type of contamination presented here has not been documented before.
Recently, \citet{batalha2012} discussed the possibility that scattered
light from bright eclipsing binaries contaminates the masks of 
neighboring targets.
However, these authors referred to targets within 20'' of the binary,
whereas the contamination that we have observed originates in 
targets more than 1 arc degree away.
The origin of this contamination deserves a more detailed study which
is beyond the scope of this paper.
The reason for this behavior might be an effect named video cross-talk
(\citealt{keplerInstrumentHandbook}; J. Jenkins, priv. comm.).

\subsubsection{False positives from the Kepler pipeline}
\label{subsubsec:nonDetectionsKeplerFalsePositives}

In addition to the cases of KOI~559.01 and KOI~712.01 described above,
there are other clear false alarms unnoticed among the planetary
candidates in \citetalias{borucki2011b}.
\begin{itemize}
\item KOI~774 shows a clear eccentric secondary eclipse on the 
  $3 \cdot 10^{-4}$ level (300 ppm), incompatible with the occultation
  of a planetary companion \citep[see][]{coughli2012}. This candidate
  has been classified as an eclipsing binary by \citet{slawson2011}.
\item KOI~823 shows out-of-transit variations compatible with a
  massive companion and a secondary eclipse on the $7 \cdot 10^{-4}$
  level, incompatible with the occultation of a planetary companion.
  This candidate was also classified as a potential false
  positive by \citet{demory2011}.
\item KOI~876 shows a secondary eclipse on the $7 \cdot 10^{-4}$
  level, incompatible with the occultation of a planetary companion.
  This candidate was also classified as a false positive by
  \citet{demory2011}.
\item KOI~960 shows an eccentric secondary eclipse on the 
  $5 \cdot 10^{-4}$ level, incompatible with the occultation of a
  planetary companion.
  This candidate was also classified as a potential false
  positive by \citet{demory2011}.
\item KOI~1285 shows a secondary eclipse on the $3 \cdot 10^{-4}$
  level, incompatible with the occultation of a planetary 
  companion. Additionally, it shows strong eclipse timing variations,
  indicating the presence of either stellar interaction or
  additional companions in the system. This candidate has been
  classified as an eclipsing binary by \citet{slawson2011}  
  and also classified as a potential false positive by
  \citet{demory2011}.
\item KOI~1401 displays the typical out-of-eclipse behavior of an
  eclipsing binary and has been classified as such by
  \citet{slawson2011}.
\item KOI~1448 is an active star, the light curve shows a secondary 
  eclipse on the $8 \cdot 10^{-4}$ level, incompatible with the
  occultation of a planetary companion, and has been classified as an
  eclipsing binary by \citet{slawson2011} 
  and also classified as a false positive by \citet{demory2011}.
\item KOI~1452 is another active star with a secondary eclipse on the
  $1 \cdot 10^{-3}$ level, incompatible with the occultation of a
  planetary companion, and has been classified as an  
  eclipsing binary by \citet{slawson2011}.
\item KOI~1459 is an eclipsing binary with a real period of 1.38 days,
  twice that found by \citetalias{borucki2011b} and
  \citet{slawson2011}. The depths of the primary and secondary
  eclipses are $0.45 \pm 0.01$ and $0.36 \pm 0.01$.
\item KOI~1541 shows a secondary eclipse on the $1 \cdot 10^{-3}$
  level, incompatible with the occultation of a planetary companion,
  and has been classified as an eclipsing binary by
  \citet{slawson2011}. 
\item KOI~1543 shows a secondary eclipse on the $1 \cdot 10^{-3}$
  level, incompatible with the occultation of a planetary
  companion. It was classified as an eclipsing binary with the wrong
  period of 7.92 days by \citet{prsa2011}, the real period is half
  that or 3.96 days, but it is not in the list of binaries by
  \citet{slawson2011} although it was classified as a false
  positive by \citet{demory2011}.
\item KOI~1546 shows out-of-transit variations compatible with an
  eclipsing binary and was classified as such by \citet{prsa2011}, but
  not by \citet{slawson2011}.
\end{itemize}

For completion, we add that the candidates 
KOI~4,     
KOI~51,    
KOI~131,   
KOI~135,   
KOI~138,  
KOI~201,  
KOI~206,  
KOI~211,  
KOI~225,  
KOI~256,  
KOI~261,  
KOI~340,  
KOI~687,  
KOI~741,  
KOI~822,  
KOI~913,  
KOI~976,  
KOI~1187, 
KOI~1020, 
KOI~1032, 
KOI~1227, and 
KOI~1385 
have also been
classified either by \citet{prsa2011} or by \citet{slawson2011} as
eclipsing binaries, and considering their light curves, they probably
are.
They all still appear as valid planetary candidates in the list of
\citet{batalha2012}, because the vetting of the first $1\,600$ KOIs
has not been reprocessed with the improved method described in that
paper which has been used for the most recent release of 
KOIs\footnote{while our paper was in the revision process,
  \citet{ofir2012} provided arguments to reclassify some additional 
  KOIs as eclipsing binaries.}.
They represent a small fraction of the total number of candidates, but
they are still readily identifiable by a careful analysis of the light
curve. 
The careful tests applied by the \kepler\,pipeline could not remove
these easily identificable false positives.
One may wonder how many of the remaining candidates with lower S/N
ratios, in which region those tests are less sensitive, will turn out
to be also false positives. 
Authors should be extremely careful when interpreting the statistical
analysis based on candidates without radial velocity confirmation or
candidates that have not been properly validated \citep{torres2011}. 



\subsubsection{Other reasons}
\label{subsubsec:nonDetectionsOtherReasons}

\begin{figure}[t]
  \begin{center}
    \includegraphics[%
      width=0.9\linewidth,%
      height=0.5\textheight,%
      keepaspectratio]{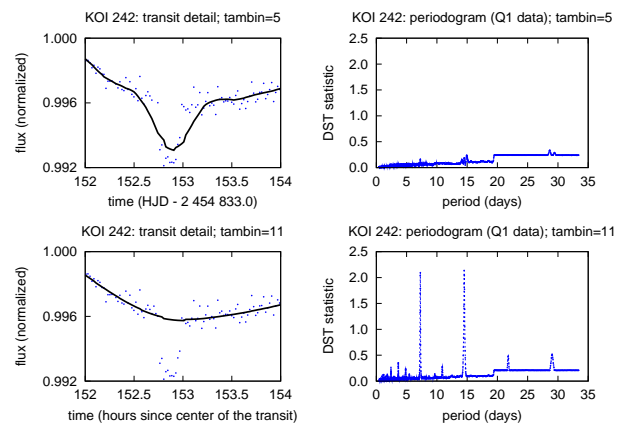}
  \end{center}
  \caption{
    Light curve of KOI~242 filtered with the original configuration
    used for the analysis in this paper (top) and with an alternative
    configuration (bottom) that prevents the filtering of long
    transits.
  }
  \label{fig:koi242}
\end{figure}

\begin{figure}[t]
  \begin{center}
    \includegraphics[%
      width=0.9\linewidth,%
      height=0.5\textheight,%
      keepaspectratio]{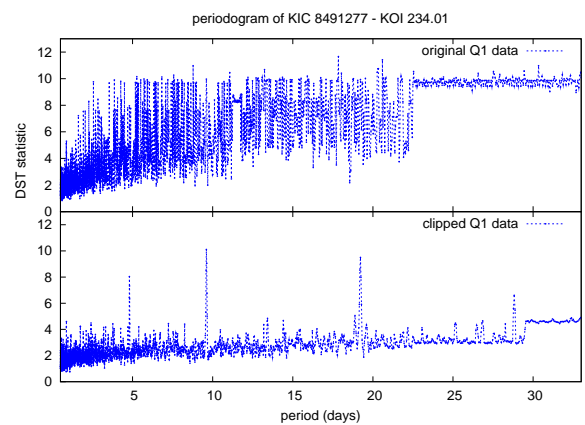}
  \end{center}
  \caption{
    Above: periodogram of KOI~234 (KIC~8491277) in the original Q1
    data set. Below, the same periodogram after applying a 4$\sigma$
    clipping to the light curve. The candidate, which could not be
    distinguished from the noise in the original data set, is clearly
    detected in the filtered data set.}
  \label{fig:kic8491277_pdg_mtp}
\end{figure}

In \noBoruckiFiltering\,cases either the stellar variability filter
failed to remove the stellar signal without damaging the transit
signal (for example, in the case of KOI~242, with a transit duration
of almost 6h, see Fig.~\ref{fig:koi242}) or some residual
outliers prevented the candidate detection.
As we mentioned above, we relied on the cosmic ray filtering of the
Kepler pipeline.
In some cases, remaining outliers in the light curve prevented the
detection of candidates such as KOI~234 (see
Fig.~\ref{fig:kic8491277_pdg_mtp}) with our algorithm,
but these cases represent a minority of the non-detections. 
The stellar variability filtering is performing well, allowing the
detection of most of the expected \kepler\,candidates, although it could
be improved to enhance the detection of weaker signals of yet undetected
transit candidates.

\subsection{False alarms from our pipeline}
\label{subsec:falseAlarms}

\begin{figure}[t]
  \begin{center}
    \includegraphics[%
      width=0.9\linewidth,%
      height=0.5\textheight,%
      keepaspectratio]{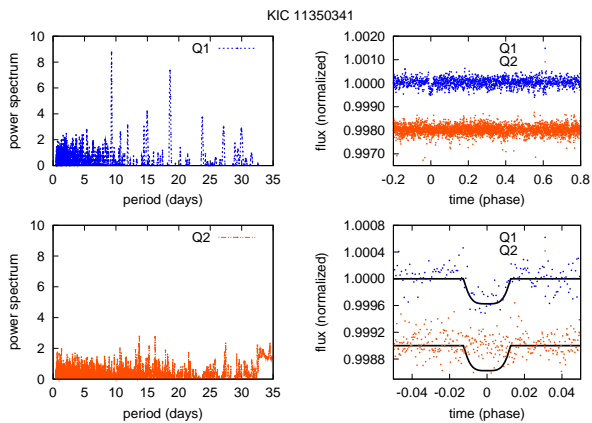}
  \end{center}
  \caption{
    Left: periodograms of the Q1 (top) and Q2 (bottom) data sets of the
    target KIC~11350341. Only the Q1 data show a periodic signal. 
    Right: filtered and folded light curve of KIC~11350341. Q1 data
    distinctly show a transiting signal, which cannot be recovered in
    Q2 data.
  }
  \label{fig:kic11350341_mtp}
\end{figure}

The false alarms in Table~\ref{table:fa} were identified because they
could be detected in the Q1 data set, but not in subsequent runs. 
We present as an illustrating example the case of KIC~11350341 (see
Fig.~\ref{fig:kic11350341_mtp}). 
The periodogram of the Q1 data set has a distinct peak at the period of
9.3 days caused by thre transit events visible in the raw light curve.
However, subsequent observations during the quarters Q2 to Q6 do not
show the presence of this periodic signal.
We do not find any nearby contaminating eclipsing binary that could
be responsible of the signal.
However, there are some nearby variable stars that contaminate the
mask.
Possibly the stellar activity filter is not completely removing
their contribution, and therefore the residuals produce the spurious
detection. 
We think that the Q1 data suffers from more contamination from
neighboring sources than subsequent runs, which could explain why
false alarms are only present in the Q1 data. 
One of the reasons for this additional contamination would be the
presence of a variable star among the guiding stars in Q1 (see the
discussion in \citealt{caldwell2010,haas2010,jenkins2010a}). 
Another possible reason are instrumental effects such as those
described above in the case of KOI~559.01.


\section{Summary}
\label{sec:summary}

We have described a new algorithm for transit detection using an
analytic model for the shape of a transit
(Section~\ref{sec:ecuaciones}). 
We have compared in Section~\ref{sec:discusion} the performance of the
new transit detection technique, \dst, with that of a widely used
technique, the BLS algorithm, and shown that our algorithm performs
better. 
We analyzed two representative test cases and we concluded
that \dst\, produces a better signal detection efficiency than
BLS due to the improved description of the transit signal and to the
improved definition of the test statistic in all relevant cases
for transit surveys: terrestrial planets, giant planets, and eclipsing
binaries. 
We discussed the advantages and the flexibility of this algorithm
in defining the region in transit
(Section~\ref{sec:paradigma}), with their implications for the search
for planets that experience significant transit timing variations
(Subsection~\ref{subsec:ttv}) and transiting multiple planet systems
(Section~\ref{sec:futuro}).
We intend to develop methods to make the computational effort required
by this algorithm more effective in the future by changing the
paradigm of the blind transit searches, as it has been described in
Section~\ref{sec:futuro}. 

We applied our algorithm and a set of tools to filter the stellar
activity, which was designed for the \corot\, search of
transiting planets, to the publicly released Q1 \kepler\,data. 
As a result of this analysis, some adjustments were made to
improve the performance of the stellar activity filtering. 
We report \noNewCandidatesBatalha\, planetary candidates, not
reported previously in the literature, based on our analysis
of 33 days of \kepler\,data.
We discussed the impact of instrumental and stellar activity
residuals in the data on the detection of planetary candidates.
Our study shows that the analysis of space mission data
advocates the use of complementary detrending and transit
detection tools for future space-based transit surveys such as
PLATO \citep{catala2009}.

%
\begin{acknowledgements}
We would like to thank Jon Jenkins and our anonymous referee for their
insightful reading of our manuscript and their detailed comments, which
improved our manuscript.
\kepler\,data presented in this paper were obtained from the
Multimission Archive at the Space Telescope Science Institute 
(MAST). STScI is operated by the Association of Universities for
Research in Astronomy, Inc., under NASA contract NAS5-26555. Support
for MAST for non-HST data is provided by the NASA Office of Space
Science via grant NNX09AF08G and by other grants and contracts 
The CoRoT space mission, launched on December 27 2006, was developed
and is operated by CNES, with contributions from  Austria, Belgium,
Brazil, ESA, Germany and Spain. 
This research has made use of NASA's Astrophysics Data System.
\end{acknowledgements}

%
\bibliographystyle{bibtex/aa}
\bibliography{bibl}

\begin{thebibliography}{77}
\expandafter\ifx\csname natexlab\endcsname\relax\def\natexlab#1{#1}\fi

\bibitem[{{Agol} {et~al.}(2005){Agol}, {Steffen}, {Sari}, \&
  {Clarkson}}]{agol2005}
{Agol}, E., {Steffen}, J., {Sari}, R., \& {Clarkson}, W. 2005, \mnras, 359, 567

\bibitem[{{Aigrain} \& {Favata}(2002)}]{aigrain2002}
{Aigrain}, S. \& {Favata}, F. 2002, \aap, 395, 625

\bibitem[{{Aigrain} {et~al.}(2009){Aigrain}, {Pont}, {Fressin}, {Alapini},
  {Alonso}, {Auvergne}, {Barbieri}, {Barge}, {Bord{\'e}}, {Bouchy}, {Deeg}, {de
  La Reza}, {Deleuil}, {Dvorak}, {Erikson}, {Fridlund}, {Gondoin}, {Guterman},
  {Jorda}, {Lammer}, {L{\'e}ger}, {Llebaria}, {Magain}, {Mazeh}, {Moutou},
  {Ollivier}, {P{\"a}tzold}, {Queloz}, {Rauer}, {Rouan}, {Schneider},
  {Wuchter}, \& {Zucker}}]{aigrain2009}
{Aigrain}, S., {Pont}, F., {Fressin}, F., {et~al.} 2009, \aap, 506, 425

\bibitem[{{Alonso} {et~al.}(2008){Alonso}, {Auvergne}, {Baglin}, {Ollivier},
  {Moutou}, {Rouan}, {Deeg}, {Aigrain}, {Almenara}, {Barbieri}, {Barge},
  {Benz}, {Bord{\'e}}, {Bouchy}, {de La Reza}, {Deleuil}, {Dvorak}, {Erikson},
  {Fridlund}, {Gillon}, {Gondoin}, {Guillot}, {Hatzes}, {H{\'e}brard},
  {Kabath}, {Jorda}, {Lammer}, {L{\'e}ger}, {Llebaria}, {Loeillet}, {Magain},
  {Mayor}, {Mazeh}, {P{\"a}tzold}, {Pepe}, {Pont}, {Queloz}, {Rauer},
  {Shporer}, {Schneider}, {Stecklum}, {Udry}, \& {Wuchterl}}]{alonso2008a}
{Alonso}, R., {Auvergne}, M., {Baglin}, A., {et~al.} 2008, \aap, 482, L21

\bibitem[{{Baglin} {et~al.}(2006){Baglin}, {Auvergne}, {Boisnard}, {Lam-Trong},
  {Barge}, {Catala}, {Deleuil}, {Michel}, \& {Weiss}}]{baglin2006}
{Baglin}, A., {Auvergne}, M., {Boisnard}, L., {et~al.} 2006, in COSPAR, Plenary
  Meeting, Vol.~36, 36th COSPAR Scientific Assembly, 3749

\bibitem[{{Barge} {et~al.}(2008){Barge}, {Baglin}, {Auvergne}, {Rauer},
  {L{\'e}ger}, {Schneider}, {Pont}, {Aigrain}, {Almenara}, {Alonso},
  {Barbieri}, {Bord{\'e}}, {Bouchy}, {Deeg}, {La Reza}, {Deleuil}, {Dvorak},
  {Erikson}, {Fridlund}, {Gillon}, {Gondoin}, {Guillot}, {Hatzes}, {Hebrard},
  {Jorda}, {Kabath}, {Lammer}, {Llebaria}, {Loeillet}, {Magain}, {Mazeh},
  {Moutou}, {Ollivier}, {P{\"a}tzold}, {Queloz}, {Rouan}, {Shporer}, \&
  {Wuchterl}}]{barge2008a}
{Barge}, P., {Baglin}, A., {Auvergne}, M., {et~al.} 2008, \aap, 482, L17

\bibitem[{{Batalha} {et~al.}(2011){Batalha}, {Borucki}, {Bryson}, {Buchhave},
  {Caldwell}, {Christensen-Dalsgaard}, {Ciardi}, {Dunham}, {Fressin},
  {Gautier}, {Gilliland}, {Haas}, {Howell}, {Jenkins}, {Kjeldsen}, {Koch},
  {Latham}, {Lissauer}, {Marcy}, {Rowe}, {Sasselov}, {Seager}, {Steffen},
  {Torres}, {Basri}, {Brown}, {Charbonneau}, {Christiansen}, {Clarke},
  {Cochran}, {Dupree}, {Fabrycky}, {Fischer}, {Ford}, {Fortney}, {Girouard},
  {Holman}, {Johnson}, {Isaacson}, {Klaus}, {Machalek}, {Moorehead},
  {Morehead}, {Ragozzine}, {Tenenbaum}, {Twicken}, {Quinn}, {VanCleve},
  {Walkowicz}, {Welsh}, {Devore}, \& {Gould}}]{batalha2011}
{Batalha}, N.~M., {Borucki}, W.~J., {Bryson}, S.~T., {et~al.} 2011, \apj, 729,
  27

\bibitem[{{Batalha} {et~al.}(2012){Batalha}, {Rowe}, {Bryson}, {Barclay},
  {Burke}, {Caldwell}, {Christiansen}, {Mullally}, {Thompson}, {Brown},
  {Dupree}, {Fabrycky}, {Ford}, {Fortney}, {Gilliland}, {Isaacson}, {Latham},
  {Marcy}, {Quinn}, {Ragozzine}, {Shporer}, {Borucki}, {Ciardi}, {Gautier},
  {Haas}, {Jenkins}, {Koch}, {Lissauer}, {Rapin}, {Basri}, {Boss}, {Buchhave},
  {Charbonneau}, {Christensen-Dalsgaard}, {Clarke}, {Cochran}, {Demory},
  {Devore}, {Esquerdo}, {Everett}, {Fressin}, {Geary}, {Girouard}, {Gould},
  {Hall}, {Holman}, {Howard}, {Howell}, {Ibrahim}, {Kinemuchi}, {Kjeldsen},
  {Klaus}, {Li}, {Lucas}, {Morris}, {Prsa}, {Quintana}, {Sanderfer},
  {Sasselov}, {Seader}, {Smith}, {Steffen}, {Still}, {Stumpe}, {Tarter},
  {Tenenbaum}, {Torres}, {Twicken}, {Uddin}, {Van Cleve}, {Walkowicz}, \&
  {Welsh}}]{batalha2012}
{Batalha}, N.~M., {Rowe}, J.~F., {Bryson}, S.~T., {et~al.} 2012, ArXiv e-prints
  1202.5852

\bibitem[{{Batalha} {et~al.}(2010){Batalha}, {Rowe}, {Gilliland}, {Jenkins},
  {Caldwell}, {Borucki}, {Koch}, {Lissauer}, {Dunham}, {Gautier}, {Howell},
  {Latham}, {Marcy}, \& {Prsa}}]{batalha2010b}
{Batalha}, N.~M., {Rowe}, J.~F., {Gilliland}, R.~L., {et~al.} 2010, \apjl, 713,
  L103

\bibitem[{{Bonomo} \& {Lanza}(2008)}]{bonomo2008}
{Bonomo}, A.~S. \& {Lanza}, A.~F. 2008, \aap, 482, 341

\bibitem[{{Bord{\'e}} {et~al.}(2007){Bord{\'e}}, {Fressin}, {Ollivier},
  {L{\'e}ger}, \& {Rouan}}]{borde2007}
{Bord{\'e}}, P., {Fressin}, F., {Ollivier}, M., {L{\'e}ger}, A., \& {Rouan}, D.
  2007, in Astronomical Society of the Pacific Conference Series, Vol. 366,
  Transiting Extrapolar Planets Workshop, ed. C.~{Afonso}, D.~{Weldrake}, \&
  T.~{Henning}, 145

\bibitem[{{Borkovits} {et~al.}(2011){Borkovits}, {Csizmadia},
  {Forg{\'a}cs-Dajka}, \& {Heged{\"u}s}}]{borkovits2011}
{Borkovits}, T., {Csizmadia}, S., {Forg{\'a}cs-Dajka}, E., \& {Heged{\"u}s}, T.
  2011, \aap, 528, A53

\bibitem[{{Borucki} {et~al.}(2010){Borucki}, {Koch}, {Basri}, {Batalha},
  {Brown}, {Caldwell}, {Caldwell}, {Christensen-Dalsgaard}, {Cochran},
  {DeVore}, {Dunham}, {Dupree}, {Gautier}, {Geary}, {Gilliland}, {Gould},
  {Howell}, {Jenkins}, {Kondo}, {Latham}, {Marcy}, {Meibom}, {Kjeldsen},
  {Lissauer}, {Monet}, {Morrison}, {Sasselov}, {Tarter}, {Boss}, {Brownlee},
  {Owen}, {Buzasi}, {Charbonneau}, {Doyle}, {Fortney}, {Ford}, {Holman},
  {Seager}, {Steffen}, {Welsh}, {Rowe}, {Anderson}, {Buchhave}, {Ciardi},
  {Walkowicz}, {Sherry}, {Horch}, {Isaacson}, {Everett}, {Fischer}, {Torres},
  {Johnson}, {Endl}, {MacQueen}, {Bryson}, {Dotson}, {Haas}, {Kolodziejczak},
  {Van Cleve}, {Chandrasekaran}, {Twicken}, {Quintana}, {Clarke}, {Allen},
  {Li}, {Wu}, {Tenenbaum}, {Verner}, {Bruhweiler}, {Barnes}, \&
  {Prsa}}]{borucki2010a}
{Borucki}, W.~J., {Koch}, D., {Basri}, G., {et~al.} 2010, Science, 327, 977

\bibitem[{{Borucki} {et~al.}(2011{\natexlab{a}}){Borucki}, {Koch}, {Basri},
  {Batalha}, {Boss}, {Brown}, {Caldwell}, {Christensen-Dalsgaard}, {Cochran},
  {DeVore}, {Dunham}, {Dupree}, {Gautier}, {Geary}, {Gilliland}, {Gould},
  {Howell}, {Jenkins}, {Kjeldsen}, {Latham}, {Lissauer}, {Marcy}, {Monet},
  {Sasselov}, {Tarter}, {Charbonneau}, {Doyle}, {Ford}, {Fortney}, {Holman},
  {Seager}, {Steffen}, {Welsh}, {Allen}, {Bryson}, {Buchhave},
  {Chandrasekaran}, {Christiansen}, {Ciardi}, {Clarke}, {Dotson}, {Endl},
  {Fischer}, {Fressin}, {Haas}, {Horch}, {Howard}, {Isaacson}, {Kolodziejczak},
  {Li}, {MacQueen}, {Meibom}, {Prsa}, {Quintana}, {Rowe}, {Sherry},
  {Tenenbaum}, {Torres}, {Twicken}, {Van Cleve}, {Walkowicz}, \&
  {Wu}}]{borucki2011a}
{Borucki}, W.~J., {Koch}, D.~G., {Basri}, G., {et~al.} 2011{\natexlab{a}},
  \apj, 728, 117

\bibitem[{{Borucki} {et~al.}(2011{\natexlab{b}}){Borucki}, {Koch}, {Basri},
  {Batalha}, {Brown}, {Bryson}, {Caldwell}, {Christensen-Dalsgaard}, {Cochran},
  {DeVore}, {Dunham}, {Gautier}, {Geary}, {Gilliland}, {Gould}, {Howell},
  {Jenkins}, {Latham}, {Lissauer}, {Marcy}, {Rowe}, {Sasselov}, {Boss},
  {Charbonneau}, {Ciardi}, {Doyle}, {Dupree}, {Ford}, {Fortney}, {Holman},
  {Seager}, {Steffen}, {Tarter}, {Welsh}, {Allen}, {Buchhave}, {Christiansen},
  {Clarke}, {Das}, {D{\'e}sert}, {Endl}, {Fabrycky}, {Fressin}, {Haas},
  {Horch}, {Howard}, {Isaacson}, {Kjeldsen}, {Kolodziejczak}, {Kulesa}, {Li},
  {Lucas}, {Machalek}, {McCarthy}, {MacQueen}, {Meibom}, {Miquel}, {Prsa},
  {Quinn}, {Quintana}, {Ragozzine}, {Sherry}, {Shporer}, {Tenenbaum}, {Torres},
  {Twicken}, {Van Cleve}, {Walkowicz}, {Witteborn}, \& {Still}}]{borucki2011b}
{Borucki}, W.~J., {Koch}, D.~G., {Basri}, G., {et~al.} 2011{\natexlab{b}},
  \apj, 736, 19

\bibitem[{{Borucki} {et~al.}(2012){Borucki}, {Koch}, {Batalha}, {Bryson},
  {Rowe}, {Fressin}, {Torres}, {Caldwell}, {Christensen-Dalsgaard}, {Cochran},
  {DeVore}, {Gautier}, {Geary}, {Gilliland}, {Gould}, {Howell}, {Jenkins},
  {Latham}, {Lissauer}, {Marcy}, {Sasselov}, {Boss}, {Charbonneau}, {Ciardi},
  {Kaltenegger}, {Doyle}, {Dupree}, {Ford}, {Fortney}, {Holman}, {Steffen},
  {Mullally}, {Still}, {Tarter}, {Ballard}, {Buchhave}, {Carter},
  {Christiansen}, {Demory}, {D{\'e}sert}, {Dressing}, {Endl}, {Fabrycky},
  {Fischer}, {Haas}, {Henze}, {Horch}, {Howard}, {Isaacson}, {Kjeldsen},
  {Johnson}, {Klaus}, {Kolodziejczak}, {Barclay}, {Li}, {Meibom}, {Prsa},
  {Quinn}, {Quintana}, {Robertson}, {Sherry}, {Shporer}, {Tenenbaum},
  {Thompson}, {Twicken}, {Van Cleve}, {Welsh}, {Basu}, {Chaplin}, {Miglio},
  {Kawaler}, {Arentoft}, {Stello}, {Metcalfe}, {Verner}, {Karoff}, {Lundkvist},
  {Lund}, {Handberg}, {Elsworth}, {Hekker}, {Huber}, {Bedding}, \&
  {Rapin}}]{borucki2012}
{Borucki}, W.~J., {Koch}, D.~G., {Batalha}, N., {et~al.} 2012, \apj, 745, 120

\bibitem[{{Cabrera} {et~al.}(2009){Cabrera}, {Fridlund}, {Ollivier},
  {Gandolfi}, {Csizmadia}, {Alonso}, {Aigrain}, {Alapini}, {Almenara}, {Barge},
  {Bonomo}, {Bord{\'e}}, {Bouchy}, {Bruntt}, {Carone}, {Carpano}, {Deeg}, {de
  La Reza}, {Deleuil}, {Dvorak}, {Erikson}, {Gillon}, {Gondoin}, {Guenther},
  {Guillot}, {Hartmann}, {Hatzes}, {Hebrard}, {Jorda}, {Lammer}, {L{\'e}ger},
  {Llebaria}, {Lovis}, {Magain}, {Mayor}, {Mazeh}, {Moutou}, {Ofir},
  {P{\"a}tzold}, {Pepe}, {Pont}, {Queloz}, {Rabus}, {Rauer}, {R{\'e}gulo},
  {Renner}, {Rouan}, {Samuel}, {Santerne}, {Schneider}, {Shporer}, {Stecklum},
  {Tingley}, {Udry}, \& {Wuchterl}}]{cabrera2009}
{Cabrera}, J., {Fridlund}, M., {Ollivier}, M., {et~al.} 2009, \aap, 506, 501

\bibitem[{{Caldwell} {et~al.}(2010){Caldwell}, {Kolodziejczak}, {Van Cleve},
  {Jenkins}, {Gazis}, {Argabright}, {Bachtell}, {Dunham}, {Geary}, {Gilliland},
  {Chandrasekaran}, {Li}, {Tenenbaum}, {Wu}, {Borucki}, {Bryson}, {Dotson},
  {Haas}, \& {Koch}}]{caldwell2010}
{Caldwell}, D.~A., {Kolodziejczak}, J.~J., {Van Cleve}, J.~E., {et~al.} 2010,
  \apjl, 713, L92

\bibitem[{{Carone} {et~al.}(2012){Carone}, {Gandolfi}, {Cabrera}, {Hatzes},
  {Deeg}, {Csizmadia}, {P{\"a}tzold}, {Weingrill}, {Aigrain}, {Alonso},
  {Alapini}, {Almenara}, {Auvergne}, {Baglin}, {Barge}, {Bonomo}, {Bord{\'e}},
  {Bouchy}, {Bruntt}, {Carpano}, {Cochran}, {Deleuil}, {D{\'{\i}}az},
  {Dreizler}, {Dvorak}, {Eisl{\"o}ffel}, {Eigm{\"u}ller}, {Endl}, {Erikson},
  {Ferraz-Mello}, {Fridlund}, {Gazzano}, {Gibson}, {Gillon}, {Gondoin},
  {Grziwa}, {G{\"u}nther}, {Guillot}, {Hartmann}, {Havel}, {H{\'e}brard},
  {Jorda}, {Kabath}, {L{\'e}ger}, {Llebaria}, {Lammer}, {Lovis}, {MacQueen},
  {Mayor}, {Mazeh}, {Moutou}, {Nortmann}, {Ofir}, {Ollivier}, {Parviainen},
  {Pepe}, {Pont}, {Queloz}, {Rabus}, {Rauer}, {R{\'e}gulo}, {Renner}, {de La
  Reza}, {Rouan}, {Santerne}, {Samuel}, {Schneider}, {Shporer}, {Stecklum},
  {Tal-Or}, {Tingley}, {Udry}, \& {Wuchterl}}]{carone2012}
{Carone}, L., {Gandolfi}, D., {Cabrera}, J., {et~al.} 2012, \aap, 538, A112

\bibitem[{{Carpano} {et~al.}(2009){Carpano}, {Cabrera}, {Alonso}, {Barge},
  {Aigrain}, {Almenara}, {Bord{\'e}}, {Bouchy}, {Carone}, {Deeg}, {de La Reza},
  {Deleuil}, {Dvorak}, {Erikson}, {Fressin}, {Fridlund}, {Gondoin}, {Guillot},
  {Hatzes}, {Jorda}, {Lammer}, {L{\'e}ger}, {Llebaria}, {Magain}, {Moutou},
  {Ofir}, {Ollivier}, {Janot-Pacheco}, {P{\"a}tzold}, {Pont}, {Queloz},
  {Rauer}, {R{\'e}gulo}, {Renner}, {Rouan}, {Samuel}, {Schneider}, \&
  {Wuchterl}}]{carpano2009}
{Carpano}, S., {Cabrera}, J., {Alonso}, R., {et~al.} 2009, \aap, 506, 491

\bibitem[{{Carpano} \& {Fridlund}(2008)}]{carpano2008}
{Carpano}, S. \& {Fridlund}, M. 2008, \aap, 485, 607

\bibitem[{{Carter} {et~al.}(2012){Carter}, {Agol}, {Chaplin}, {Basu},
  {Bedding}, {Buchhave}, {Christensen-Dalsgaard}, {Deck}, {Elsworth},
  {Fabrycky}, {Ford}, {Fortney}, {Hale}, {Handberg}, {Hekker}, {Holman},
  {Huber}, {Karoff}, {Kawaler}, {Kjeldsen}, {Lissauer}, {Lopez}, {Lund},
  {Lundkvist}, {Metcalfe}, {Miglio}, {Rogers}, {Stello}, {Borucki}, {Bryson},
  {Christiansen}, {Cochran}, {Geary}, {Gilliland}, {Haas}, {Hall}, {Howard},
  {Jenkins}, {Klaus}, {Koch}, {Latham}, {MacQueen}, {Sasselov}, {Steffen},
  {Twicken}, \& {Winn}}]{carter2012}
{Carter}, J.~A., {Agol}, E., {Chaplin}, W.~J., {et~al.} 2012, Science, 337, 556

\bibitem[{{Catala}(2009)}]{catala2009}
{Catala}, C. 2009, Experimental Astronomy, 23, 329

\bibitem[{{Cavarroc} {et~al.}(2012){Cavarroc}, {Moutou}, {Gandolfi}, {Tingley},
  {Ollivier}, {Aigrain}, {Alonso}, {Almenara}, {Auvergne}, {Baglin}, {Barge},
  {Bonomo}, {Bord{\'e}}, {Bouchy}, {Cabrera}, {Carpano}, {Carone}, {Cochran},
  {Csizmadia}, {Deeg}, {Deleuil}, {D{\'{\i}}az}, {Dvorak}, {Endl}, {Erikson},
  {Fridlund}, {Gillon}, {Guenther}, {Guillot}, {Hatzes}, {H{\'e}brard},
  {Jorda}, {L{\'e}ger}, {Lammer}, {Lev}, {Lovis}, {MacQueen}, {Mazeh}, {Ofir},
  {Parviainen}, {Pasternacki}, {P{\"a}tzold}, {Queloz}, {Rauer}, {Rouan},
  {Samuel}, {Santerne}, {Schneider}, {Weingrill}, \& {Wuchterl}}]{cavarroc2012}
{Cavarroc}, C., {Moutou}, C., {Gandolfi}, D., {et~al.} 2012, \apss, 337, 511

\bibitem[{{Charbonneau} {et~al.}(2009){Charbonneau}, {Berta}, {Irwin}, {Burke},
  {Nutzman}, {Buchhave}, {Lovis}, {Bonfils}, {Latham}, {Udry}, {Murray-Clay},
  {Holman}, {Falco}, {Winn}, {Queloz}, {Pepe}, {Mayor}, {Delfosse}, \&
  {Forveille}}]{charbonneau2009}
{Charbonneau}, D., {Berta}, Z.~K., {Irwin}, J., {et~al.} 2009, \nat, 462, 891

\bibitem[{{Clarkson} {et~al.}(2007){Clarkson}, {Enoch}, {Haswell}, {Norton},
  {Christian}, {Collier Cameron}, {Kane}, {Horne}, {Lister}, {Street}, {West},
  {Wilson}, {Evans}, {Fitzsimmons}, {Hellier}, {Hodgkin}, {Irwin}, {Keenan},
  {Osborne}, {Parley}, {Pollacco}, {Ryans}, {Skillen}, \&
  {Wheatley}}]{clarkson2007}
{Clarkson}, W.~I., {Enoch}, B., {Haswell}, C.~A., {et~al.} 2007, \mnras, 381,
  851

\bibitem[{{Collier Cameron} {et~al.}(2006){Collier Cameron}, {Pollacco},
  {Street}, {Lister}, {West}, {Wilson}, {Pont}, {Christian}, {Clarkson},
  {Enoch}, {Evans}, {Fitzsimmons}, {Haswell}, {Hellier}, {Hodgkin}, {Horne},
  {Irwin}, {Kane}, {Keenan}, {Norton}, {Parley}, {Osborne}, {Ryans}, {Skillen},
  \& {Wheatley}}]{colliercameron2006}
{Collier Cameron}, A., {Pollacco}, D., {Street}, R.~A., {et~al.} 2006, \mnras,
  373, 799

\bibitem[{{Coughlin} \& {L{\'o}pez-Morales}(2012)}]{coughli2012}
{Coughlin}, J.~L. \& {L{\'o}pez-Morales}, M. 2012, \aj, 143, 39

\bibitem[{{Croll} {et~al.}(2007){Croll}, {Matthews}, {Rowe}, {Gladman},
  {Miller-Ricci}, {Sasselov}, {Walker}, {Kuschnig}, {Lin}, {Guenther},
  {Moffat}, {Rucinski}, \& {Weiss}}]{croll2007}
{Croll}, B., {Matthews}, J.~M., {Rowe}, J.~F., {et~al.} 2007, \apj, 671, 2129

\bibitem[{{Dawson} {et~al.}(2012){Dawson}, {Johnson}, {Morton}, {Crepp},
  {Fabrycky}, {Murray-Clay}, \& {Howard}}]{dawson2012}
{Dawson}, R.~I., {Johnson}, J.~A., {Morton}, T.~D., {et~al.} 2012, ArXiv
  e-prints 1206.5579

\bibitem[{{Deeg} {et~al.}(2009){Deeg}, {Gillon}, {Shporer}, {Rouan},
  {Stecklum}, {Aigrain}, {Alapini}, {Almenara}, {Alonso}, {Barbieri}, {Bouchy},
  {Eisl{\"o}ffel}, {Erikson}, {Fridlund}, {Eigm{\"u}ller}, {Handler}, {Hatzes},
  {Kabath}, {Lendl}, {Mazeh}, {Moutou}, {Queloz}, {Rauer}, {Rabus}, {Tingley},
  \& {Titz}}]{deeg2009}
{Deeg}, H.~J., {Gillon}, M., {Shporer}, A., {et~al.} 2009, \aap, 506, 343

\bibitem[{{Deeg} {et~al.}(2010){Deeg}, {Moutou}, {Erikson}, {Csizmadia},
  {Tingley}, {Barge}, {Bruntt}, {Havel}, {Aigrain}, {Almenara}, {Alonso},
  {Auvergne}, {Baglin}, {Barbieri}, {Benz}, {Bonomo}, {Bord{\'e}}, {Bouchy},
  {Cabrera}, {Carone}, {Carpano}, {Ciardi}, {Deleuil}, {Dvorak},
  {Ferraz-Mello}, {Fridlund}, {Gandolfi}, {Gazzano}, {Gillon}, {Gondoin},
  {Guenther}, {Guillot}, {Hartog}, {Hatzes}, {Hidas}, {H{\'e}brard}, {Jorda},
  {Kabath}, {Lammer}, {L{\'e}ger}, {Lister}, {Llebaria}, {Lovis}, {Mayor},
  {Mazeh}, {Ollivier}, {P{\"a}tzold}, {Pepe}, {Pont}, {Queloz}, {Rabus},
  {Rauer}, {Rouan}, {Samuel}, {Schneider}, {Shporer}, {Stecklum}, {Street},
  {Udry}, {Weingrill}, \& {Wuchterl}}]{deeg2010}
{Deeg}, H.~J., {Moutou}, C., {Erikson}, A., {et~al.} 2010, \nat, 464, 384

\bibitem[{{Defa{\"y}} {et~al.}(2001){Defa{\"y}}, {Deleuil}, \&
  {Barge}}]{defay2001}
{Defa{\"y}}, C., {Deleuil}, M., \& {Barge}, P. 2001, \aap, 365, 330

\bibitem[{{Deming} {et~al.}(2011){Deming}, {Sada}, {Jackson}, {Peterson},
  {Agol}, {Knutson}, {Jennings}, {Haase}, \& {Bays}}]{deming2011}
{Deming}, D., {Sada}, P.~V., {Jackson}, B., {et~al.} 2011, \apj, 740, 33

\bibitem[{{Demory} \& {Seager}(2011)}]{demory2011}
{Demory}, B.-O. \& {Seager}, S. 2011, \apjs, 197, 12

\bibitem[{{Erikson} {et~al.}(2012){Erikson}, {Santerne}, {Renner}, {Barge},
  {Aigrain}, {Alapini}, {Almenara}, {Alonso}, {Auvergne}, {Baglin}, {Benz},
  {Bonomo}, {Bord{\'e}}, {Bouchy}, {Bruntt}, {Cabrera}, {Carone}, {Carpano},
  {Csizmadia}, {Deleuil}, {Deeg}, {D{\'{\i}}az}, {Dvorak}, {Ferraz-Mello},
  {Fridlund}, {Gandolfi}, {Gazzano}, {Gillon}, {Guenther}, {Guillot}, {Hatzes},
  {H{\'e}brard}, {Jorda}, {Lammer}, {L{\'e}ger}, {Llebaria}, {Mayor}, {Mazeh},
  {Moutou}, {Ollivier}, {Ofir}, {P{\"a}tzold}, {Pepe}, {Pont}, {Queloz},
  {Rabus}, {Rauer}, {R{\'e}gulo}, {Rouan}, {Samuel}, {Schneider}, {Shporer},
  {Tingley}, {Udry}, \& {Wuchterl}}]{erikson2012}
{Erikson}, A., {Santerne}, A., {Renner}, S., {et~al.} 2012, \aap, 539, A14

\bibitem[{{Faedi} {et~al.}(2011){Faedi}, {West}, {Burleigh}, {Goad}, \&
  {Hebb}}]{faedi2011}
{Faedi}, F., {West}, R.~G., {Burleigh}, M.~R., {Goad}, M.~R., \& {Hebb}, L.
  2011, \mnras, 410, 899

\bibitem[{{Gilliland} {et~al.}(2011){Gilliland}, {Chaplin}, {Dunham},
  {Argabright}, {Borucki}, {Basri}, {Bryson}, {Buzasi}, {Caldwell}, {Elsworth},
  {Jenkins}, {Koch}, {Kolodziejczak}, {Miglio}, {van Cleve}, {Walkowicz}, \&
  {Welsh}}]{gilliland2011}
{Gilliland}, R.~L., {Chaplin}, W.~J., {Dunham}, E.~W., {et~al.} 2011, \apjs,
  197, 6

\bibitem[{{Grziwa} {et~al.}(2012){Grziwa}, {P{\"a}tzold}, \&
  {Carone}}]{grziwa2012}
{Grziwa}, S., {P{\"a}tzold}, M., \& {Carone}, L. 2012, \mnras, 420, 1045

\bibitem[{{Haas} {et~al.}(2010){Haas}, {Batalha}, {Bryson}, {Caldwell},
  {Dotson}, {Hall}, {Jenkins}, {Klaus}, {Koch}, {Kolodziejczak}, {Middour},
  {Smith}, {Sobeck}, {Stober}, {Thompson}, \& {Van Cleve}}]{haas2010}
{Haas}, M.~R., {Batalha}, N.~M., {Bryson}, S.~T., {et~al.} 2010, \apjl, 713,
  L115

\bibitem[{{Hebb} {et~al.}(2010){Hebb}, {Collier-Cameron}, {Triaud}, {Lister},
  {Smalley}, {Maxted}, {Hellier}, {Anderson}, {Pollacco}, {Gillon}, {Queloz},
  {West}, {Bentley}, {Enoch}, {Haswell}, {Horne}, {Mayor}, {Pepe}, {Segransan},
  {Skillen}, {Udry}, \& {Wheatley}}]{hebb2010}
{Hebb}, L., {Collier-Cameron}, A., {Triaud}, A.~H.~M.~J., {et~al.} 2010, \apj,
  708, 224

\bibitem[{{Hellier} {et~al.}(2009){Hellier}, {Anderson}, {Collier Cameron},
  {Gillon}, {Hebb}, {Maxted}, {Queloz}, {Smalley}, {Triaud}, {West}, {Wilson},
  {Bentley}, {Enoch}, {Horne}, {Irwin}, {Lister}, {Mayor}, {Parley}, {Pepe},
  {Pollacco}, {Segransan}, {Udry}, \& {Wheatley}}]{hellier2009}
{Hellier}, C., {Anderson}, D.~R., {Collier Cameron}, A., {et~al.} 2009, \nat,
  460, 1098

\bibitem[{{Hellier} {et~al.}(2011){Hellier}, {Anderson}, {Collier Cameron},
  {Gillon}, {Jehin}, {Lendl}, {Maxted}, {Pepe}, {Pollacco}, {Queloz},
  {S{\'e}gransan}, {Smalley}, {Smith}, {Southworth}, {Triaud}, {Udry}, \&
  {West}}]{hellier2011}
{Hellier}, C., {Anderson}, D.~R., {Collier Cameron}, A., {et~al.} 2011, \aap,
  535, L7

\bibitem[{{Holman} {et~al.}(2010){Holman}, {Fabrycky}, {Ragozzine}, {Ford},
  {Steffen}, {Welsh}, {Lissauer}, {Latham}, {Marcy}, {Walkowicz}, {Batalha},
  {Jenkins}, {Rowe}, {Cochran}, {Fressin}, {Torres}, {Buchhave}, {Sasselov},
  {Borucki}, {Koch}, {Basri}, {Brown}, {Caldwell}, {Charbonneau}, {Dunham},
  {Gautier}, {Geary}, {Gilliland}, {Haas}, {Howell}, {Ciardi}, {Endl},
  {Fischer}, {F{\"u}r{\'e}sz}, {Hartman}, {Isaacson}, {Johnson}, {MacQueen},
  {Moorhead}, {Morehead}, \& {Orosz}}]{holman2010}
{Holman}, M.~J., {Fabrycky}, D.~C., {Ragozzine}, D., {et~al.} 2010, Science,
  330, 51

\bibitem[{{Holman} \& {Murray}(2005)}]{holman2005}
{Holman}, M.~J. \& {Murray}, N.~W. 2005, Science, 307, 1288

\bibitem[{{Jenkins}(2002)}]{jenkins2002a}
{Jenkins}, J.~M. 2002, \apj, 575, 493

\bibitem[{{Jenkins} {et~al.}(2002){Jenkins}, {Caldwell}, \&
  {Borucki}}]{jenkins2002b}
{Jenkins}, J.~M., {Caldwell}, D.~A., \& {Borucki}, W.~J. 2002, \apj, 564, 495

\bibitem[{{Jenkins} {et~al.}(2010{\natexlab{a}}){Jenkins}, {Caldwell},
  {Chandrasekaran}, {Twicken}, {Bryson}, {Quintana}, {Clarke}, {Li}, {Allen},
  {Tenenbaum}, {Wu}, {Klaus}, {Van Cleve}, {Dotson}, {Haas}, {Gilliland},
  {Koch}, \& {Borucki}}]{jenkins2010a}
{Jenkins}, J.~M., {Caldwell}, D.~A., {Chandrasekaran}, H., {et~al.}
  2010{\natexlab{a}}, \apjl, 713, L120

\bibitem[{{Jenkins} {et~al.}(2010{\natexlab{b}}){Jenkins}, {Chandrasekaran},
  {McCauliff}, {Caldwell}, {Tenenbaum}, {Li}, {Klaus}, {Cote}, \&
  {Middour}}]{jenkins2010c}
{Jenkins}, J.~M., {Chandrasekaran}, H., {McCauliff}, S.~D., {et~al.}
  2010{\natexlab{b}}, in Presented at the Society of Photo-Optical
  Instrumentation Engineers (SPIE) Conference, Vol. 7740, Society of
  Photo-Optical Instrumentation Engineers (SPIE) Conference Series

\bibitem[{{Jenkins} {et~al.}(1996){Jenkins}, {Doyle}, \&
  {Cullers}}]{jenkins1996}
{Jenkins}, J.~M., {Doyle}, L.~R., \& {Cullers}, D.~K. 1996, \icarus, 119, 244

\bibitem[{{Kov{\'a}cs} {et~al.}(2002){Kov{\'a}cs}, {Zucker}, \&
  {Mazeh}}]{kovacs2002}
{Kov{\'a}cs}, G., {Zucker}, S., \& {Mazeh}, T. 2002, \aap, 391, 369

\bibitem[{{L{\'e}ger} {et~al.}(2009){L{\'e}ger}, {Rouan}, {Schneider}, {Barge},
  {Fridlund}, {Samuel}, {Ollivier}, {Guenther}, {Deleuil}, {Deeg}, {Auvergne},
  {Alonso}, {Aigrain}, {Alapini}, {Almenara}, {Baglin}, {Barbieri}, {Bruntt},
  {Bord{\'e}}, {Bouchy}, {Cabrera}, {Catala}, {Carone}, {Carpano}, {Csizmadia},
  {Dvorak}, {Erikson}, {Ferraz-Mello}, {Foing}, {Fressin}, {Gandolfi},
  {Gillon}, {Gondoin}, {Grasset}, {Guillot}, {Hatzes}, {H{\'e}brard}, {Jorda},
  {Lammer}, {Llebaria}, {Loeillet}, {Mayor}, {Mazeh}, {Moutou}, {P{\"a}tzold},
  {Pont}, {Queloz}, {Rauer}, {Renner}, {Samadi}, {Shporer}, {Sotin}, {Tingley},
  {Wuchterl}, {Adda}, {Agogu}, {Appourchaux}, {Ballans}, {Baron}, {Beaufort},
  {Bellenger}, {Berlin}, {Bernardi}, {Blouin}, {Baudin}, {Bodin}, {Boisnard},
  {Boit}, {Bonneau}, {Borzeix}, {Briet}, {Buey}, {Butler}, {Cailleau},
  {Cautain}, {Chabaud}, {Chaintreuil}, {Chiavassa}, {Costes}, {Cuna Parrho},
  {de Oliveira Fialho}, {Decaudin}, {Defise}, {Djalal}, {Epstein}, {Exil},
  {Faur{\'e}}, {Fenouillet}, {Gaboriaud}, {Gallic}, {Gamet}, {Gavalda},
  {Grolleau}, {Gruneisen}, {Gueguen}, {Guis}, {Guivarc'h}, {Guterman},
  {Hallouard}, {Hasiba}, {Heuripeau}, {Huntzinger}, {Hustaix}, {Imad},
  {Imbert}, {Johlander}, {Jouret}, {Journoud}, {Karioty}, {Kerjean},
  {Lafaille}, {Lafond}, {Lam-Trong}, {Landiech}, {Lapeyrere}, {Larqu{\'e}},
  {Laudet}, {Lautier}, {Lecann}, {Lefevre}, {Leruyet}, {Levacher}, {Magnan},
  {Mazy}, {Mertens}, {Mesnager}, {Meunier}, {Michel}, {Monjoin}, {Naudet},
  {Nguyen-Kim}, {Orcesi}, {Ottacher}, {Perez}, {Peter}, {Plasson}, {Plesseria},
  {Pontet}, {Pradines}, {Quentin}, {Reynaud}, {Rolland}, {Rollenhagen},
  {Romagnan}, {Russ}, {Schmidt}, {Schwartz}, {Sebbag}, {Sedes}, {Smit},
  {Steller}, {Sunter}, {Surace}, {Tello}, {Tiph{\`e}ne}, {Toulouse}, {Ulmer},
  {Vandermarcq}, {Vergnault}, {Vuillemin}, \& {Zanatta}}]{leger2009}
{L{\'e}ger}, A., {Rouan}, D., {Schneider}, J., {et~al.} 2009, \aap, 506, 287

\bibitem[{{Mandel} \& {Agol}(2002)}]{mandel2002}
{Mandel}, K. \& {Agol}, E. 2002, \apjl, 580, L171

\bibitem[{{Moutou} {et~al.}(2007){Moutou}, {Aigrain}, {Almenara}, {Alonso},
  {Auvergne}, {Barge}, {Blouin}, {Borde}, {Cabrera}, {Carone}, {Cautain},
  {Deeg}, {Erikson}, {Fressin}, {Guis}, {Leger}, {Guterman}, {Irwin}, {Kabath},
  {Lanza}, {Maceroni}, {Mazeh}, {Ollivier}, {Pont}, {Paetzold}, {Queloz},
  {Rauer}, {Rouan}, {Schneider}, {Tamuz}, {Voss}, \& {Zucker}}]{moutou2007}
{Moutou}, C., {Aigrain}, S., {Almenara}, J., {et~al.} 2007, in Astronomical
  Society of the Pacific Conference Series, Vol. 366, Transiting Extrapolar
  Planets Workshop, ed. C.~{Afonso}, D.~{Weldrake}, \& T.~{Henning}, 127--132

\bibitem[{{Moutou} {et~al.}(2005){Moutou}, {Pont}, {Barge}, {Aigrain},
  {Auvergne}, {Blouin}, {Cautain}, {Erikson}, {Guis}, {Guterman}, {Irwin},
  {Lanza}, {Queloz}, {Rauer}, {Voss}, \& {Zucker}}]{moutou2005}
{Moutou}, C., {Pont}, F., {Barge}, P., {et~al.} 2005, \aap, 437, 355

\bibitem[{{Ofir}(2008)}]{ofir2008}
{Ofir}, A. 2008, \mnras, 387, 1597

\bibitem[{{Ofir} \& {Dreizler}(2012)}]{ofir2012}
{Ofir}, A. \& {Dreizler}, S. 2012, ArXiv e-prints 1206.5347

\bibitem[{{Pont} {et~al.}(2006){Pont}, {Zucker}, \& {Queloz}}]{pont2006}
{Pont}, F., {Zucker}, S., \& {Queloz}, D. 2006, \mnras, 373, 231

\bibitem[{{Press} {et~al.}(2002){Press}, {Teukolsky}, {Vetterling}, \&
  {Flannery}}]{press2002}
{Press}, W.~H., {Teukolsky}, S.~A., {Vetterling}, W.~T., \& {Flannery}, B.~P.
  2002, Numerical Recipes in C++, 2nd edn. (Cambridge University Press)

\bibitem[{{Protopapas} {et~al.}(2005){Protopapas}, {Jimenez}, \&
  {Alcock}}]{protopapas2005}
{Protopapas}, P., {Jimenez}, R., \& {Alcock}, C. 2005, \mnras, 362, 460

\bibitem[{{Pr{\v s}a} {et~al.}(2011){Pr{\v s}a}, {Batalha}, {Slawson}, {Doyle},
  {Welsh}, {Orosz}, {Seager}, {Rucker}, {Mjaseth}, {Engle}, {Conroy},
  {Jenkins}, {Caldwell}, {Koch}, \& {Borucki}}]{prsa2011}
{Pr{\v s}a}, A., {Batalha}, N., {Slawson}, R.~W., {et~al.} 2011, \aj, 141, 83

\bibitem[{{Queloz} {et~al.}(2009){Queloz}, {Bouchy}, {Moutou}, {Hatzes},
  {H{\'e}brard}, {Alonso}, {Auvergne}, {Baglin}, {Barbieri}, {Barge}, {Benz},
  {Bord{\'e}}, {Deeg}, {Deleuil}, {Dvorak}, {Erikson}, {Ferraz Mello},
  {Fridlund}, {Gandolfi}, {Gillon}, {Guenther}, {Guillot}, {Jorda}, {Hartmann},
  {Lammer}, {L{\'e}ger}, {Llebaria}, {Lovis}, {Magain}, {Mayor}, {Mazeh},
  {Ollivier}, {P{\"a}tzold}, {Pepe}, {Rauer}, {Rouan}, {Schneider},
  {Segransan}, {Udry}, \& {Wuchterl}}]{queloz2009}
{Queloz}, D., {Bouchy}, F., {Moutou}, C., {et~al.} 2009, \aap, 506, 303

\bibitem[{{R{\'e}gulo} {et~al.}(2007){R{\'e}gulo}, {Almenara}, {Alonso},
  {Deeg}, \& {Roca Cort{\'e}s}}]{regulo2007}
{R{\'e}gulo}, C., {Almenara}, J.~M., {Alonso}, R., {Deeg}, H., \& {Roca
  Cort{\'e}s}, T. 2007, \aap, 467, 1345

\bibitem[{{Renner} {et~al.}(2008){Renner}, {Rauer}, {Erikson}, {Hedelt},
  {Kabath}, {Titz}, \& {Voss}}]{renner2008}
{Renner}, S., {Rauer}, H., {Erikson}, A., {et~al.} 2008, \aap, 492, 617

\bibitem[{{Sanchis-Ojeda} \& {Winn}(2011)}]{sanchisojeda2011}
{Sanchis-Ojeda}, R. \& {Winn}, J.~N. 2011, \apj, 743, 61

\bibitem[{{Savitzky} \& {Golay}(1964)}]{savitzky1964}
{Savitzky}, A. \& {Golay}, M.~J.~E. 1964, Analytical Chemistry, 36, 1627

\bibitem[{{Scargle}(1982)}]{scargle1982}
{Scargle}, J.~D. 1982, \apj, 263, 835

\bibitem[{{Schwarzenberg-Czerny}(1989)}]{schwarzenberg-czerny1989}
{Schwarzenberg-Czerny}, A. 1989, \mnras, 241, 153

\bibitem[{{Schwarzenberg-Czerny}(1998)}]{schwarzenberg-czerny1998}
{Schwarzenberg-Czerny}, A. 1998, Baltic Astronomy, 7, 43

\bibitem[{{Schwarzenberg-Czerny} \&
  {Beaulieu}(2006)}]{schwarzenberg-czerny2006}
{Schwarzenberg-Czerny}, A. \& {Beaulieu}, J. 2006, \mnras, 365, 165

\bibitem[{{Seager} \& {Mall{\'e}n-Ornelas}(2003)}]{seager2003}
{Seager}, S. \& {Mall{\'e}n-Ornelas}, G. 2003, \apj, 585, 1038

\bibitem[{{Slawson} {et~al.}(2011){Slawson}, {Pr{\v s}a}, {Welsh}, {Orosz},
  {Rucker}, {Batalha}, {Doyle}, {Engle}, {Conroy}, {Coughlin}, {Gregg},
  {Fetherolf}, {Short}, {Windmiller}, {Fabrycky}, {Howell}, {Jenkins}, {Uddin},
  {Mullally}, {Seader}, {Thompson}, {Sanderfer}, {Borucki}, \&
  {Koch}}]{slawson2011}
{Slawson}, R.~W., {Pr{\v s}a}, A., {Welsh}, W.~F., {et~al.} 2011, \aj, 142, 160

\bibitem[{{Tingley}(2003{\natexlab{a}})}]{tingley2003a}
{Tingley}, B. 2003{\natexlab{a}}, \aap, 403, 329

\bibitem[{{Tingley}(2003{\natexlab{b}})}]{tingley2003b}
{Tingley}, B. 2003{\natexlab{b}}, \aap, 408, L5

\bibitem[{{Torres} {et~al.}(2011){Torres}, {Fressin}, {Batalha}, {Borucki},
  {Brown}, {Bryson}, {Buchhave}, {Charbonneau}, {Ciardi}, {Dunham}, {Fabrycky},
  {Ford}, {Gautier}, {Gilliland}, {Holman}, {Howell}, {Isaacson}, {Jenkins},
  {Koch}, {Latham}, {Lissauer}, {Marcy}, {Monet}, {Prsa}, {Quinn}, {Ragozzine},
  {Rowe}, {Sasselov}, {Steffen}, \& {Welsh}}]{torres2011}
{Torres}, G., {Fressin}, F., {Batalha}, N.~M., {et~al.} 2011, \apj, 727, 24

\bibitem[{{Van Cleve} \& {Caldwell}(2009)}]{keplerInstrumentHandbook}
{Van Cleve}, J.~E. \& {Caldwell}, D.~A. 2009, Kepler Instrument Handbook
  (KSCI-19033) (NASA)

\bibitem[{{Winn} {et~al.}(2011){Winn}, {Matthews}, {Dawson}, {Fabrycky},
  {Holman}, {Kallinger}, {Kuschnig}, {Sasselov}, {Dragomir}, {Guenther},
  {Moffat}, {Rowe}, {Rucinski}, \& {Weiss}}]{winn2011a}
{Winn}, J.~N., {Matthews}, J.~M., {Dawson}, R.~I., {et~al.} 2011, \apjl, 737,
  L18

\end{thebibliography}

\longtabL{3}{
  \begin{landscape}
    \renewcommand{\arraystretch}{1.2}
    \begin{longtable}{*{5}{c}*{4}{r@{$\pm$}l}*{3}{c}l}
      \caption{\label{table:candidates}Parameters from the planetary candidates.} \\
      \hline\hline
      KIC & KOI & rank & SNR & Kp  & \multicolumn{2}{c}{period} & \multicolumn{2}{c}{epoch}               & \multicolumn{2}{c}{duration} & \multicolumn{2}{c}{depth} & $R_s$       & $T_\mathrm{eff}$ & $\log g$ & comments \\
          &     &      &     & mag & \multicolumn{2}{c}{[days]} & \multicolumn{2}{c}{[HJD-$2\,454\,900$]} & \multicolumn{2}{c}{[h]}      & \multicolumn{2}{c}{[ppm]} & $[R_\cdot]$ & [K]              & [cgs]    &          \\ 
      \hline
      \endfirsthead
      \caption{continued.}\\
      \hline\hline
      KIC & KOI & rank & SNR & Kp  & \multicolumn{2}{c}{period} & \multicolumn{2}{c}{epoch}               & \multicolumn{2}{c}{duration} & \multicolumn{2}{c}{depth} & $R_s$       & $T_\mathrm{eff}$ & $\log g$ & comments \\
          &     &      &     & mag & \multicolumn{2}{c}{[days]} & \multicolumn{2}{c}{[HJD-$2\,454\,900$]} & \multicolumn{2}{c}{[h]}      & \multicolumn{2}{c}{[ppm]} & $[R_\cdot]$ & [K]              & [cgs]    &          \\ 
      \hline
      \endhead
      \hline
      \endfoot
      3128552.01  & 2055.01 & 3 & 10 & 14.523 &     8.6822 &     0.0011 &   109.4164 &     0.0060 &      3.3 & 0.2 &  500 &   60 &  0.76 & $5\,530$ & 4.67 &                                                                \\ 
      4144576.01  & 2202.01 & 3 & 12 & 14.113 &   0.813146 &   0.000002 &    64.8261 &     0.0008 &      1.5 & 0.1 &  160 &   10 &  0.74 & $5\,123$ & 4.66 & CoRoT 7b-like                                                  \\
      4263293.01  & 1895.01 & 2 &  8 & 15.862 &    8.45936 &    0.00074 &   109.1483 &     0.0040 &      2.5 & 0.2 & 1300 &  170 &  0.56 & $4\,288$ & 4.73 &                                                                \\
      4263529.01  &       - & 2 &  9 & 14.031 &   10.10243 &    0.00094 &   106.3196 &     0.0058 &      1.5 & 0.5 & 1000 &  100 &  0.89 & $5\,330$ & 4.53 &                                                                \\
      4278221.01  & 1615.01 & 2 & 12 & 11.524 &   1.340907 &   0.000042 &   105.4496 &     0.0013 &      1.5 & 0.1 &  120 &   10 &  0.88 & $5\,534$ & 4.56 & CoRoT 7b-like                                                  \\
      4644952.01  & 1805.01 & 2 &  8 & 13.828 &    6.94130 &    0.00052 &   112.1919 &     0.0039 &      1.6 & 0.1 &  890 &   60 &  1.62 & $5\,647$ & 4.08 &                                                                \\
      4917596.01  & 1973.01 & 2 & 10 & 15.753 &    3.29063 &    0.00023 &   105.6227 &     0.0020 &      1.3 & 0.1 &  780 &  120 &  0.79 & $4\,360$ & 5.49 &                                                                \\ 
      5080636.01  & 1843.01 & 2 & 12 & 14.404 &    4.19411 &    0.00014 &   109.0747 &     0.0014 &      1.3 & 0.1 &  820 &   70 &  0.63 & $3\,673$ & 4.51 &                                                                \\
      5299459.02  & 1576.01 & 1 & 13 & 14.072 &    13.0843 &     0.0010 &    65.6296 &     0.0023 &      3.1 & 0.1 &  600 &   20 &  0.91 & $5\,214$ & 4.51 &                                                                \\
      5369827.01  &       - & 2 & 11 & 13.461 &    3.60718 &    0.00013 &   107.4454 &     0.0016 &      1.2 & 0.1 &  350 &   30 &  1.16 & $5\,367$ & 4.33 &                                                                \\
      5371776.02  & 1557.03 & 1 & 25 & 14.840 &    5.31590 &    0.00016 &    65.4309 &     0.0008 &      1.3 & 0.1 &  850 &   40 &  0.98 & $5\,966$ & 4.49 &                                                                \\
      5371776.03  & 1557.02 & 1 & 23 & 14.840 &    9.65335 &    0.00008 &    67.8572 &     0.0019 &      2.9 & 0.1 & 1200 &   30 &  0.98 & $5\,966$ & 4.49 &                                                                \\
      5387843.01  & 1762.01 & 2 & 12 & 14.996 &   0.870706 &   0.000002 &    65.0811 &     0.0008 &      1.2 & 0.1 &  390 &   10 &  0.93 & $5\,726$ & 4.52 &                                                                \\
      5443775.01  &       - & 2 & 15 & 12.936 &    3.30751 &    0.00016 &   107.5662 &     0.0026 &      1.9 & 0.1 &  260 &   20 &  1.10 & $5\,782$ & 4.39 &                                                                \\
      5794570.01  &       - & 1 & 14 & 12.430 &    5.44832 &    0.00025 &   109.0765 &     0.0021 &      2.6 & 0.1 &  530 &   20 &  0.56 & $5\,479$ & 4.92 & active star, $P_\mathrm{rot} \sim 6$d; peak to peak 3\%.       \\
      6153672.01  &       - & 3 & 17 & 15.337 &    3.36477 &    0.00010 &   108.4011 &     0.0010 &      1.3 & 0.1 & 3300 &  200 &  0.85 & $5\,600$ & 4.59 & pulsating bg star with period 0.31d                            \\ 
      6616218.01  & 1692.01 & 1 &  - & 12.557 &    2.46112 &    0.00030 &    66.3108 &     0.0040 &      1.6 & 0.1 &   66 &   10 &  0.96 & $5\,224$ & 4.47 & \\
      6616218.02  & 1692.02 & 1 & 15 & 12.557 &    5.96039 &    0.00002 &    70.2641 &     0.0010 &      2.9 & 0.1 &  720 &   10 &  0.96 & $5\,224$ & 4.47 &                                                                \\
      6696462.01  & 1693.01 & 3 &  9 & 14.980 &    3.40377 &    0.00010 &   107.6720 &     0.0032 &      1.2 & 0.1 &  530 &   70 &  0.89 & $5\,384$ & 4.54 &                                                                \\ 
      7202957.01  &       - & 2 &  9 & 10.158 &    1.71674 &    0.00007 &   106.2430 &     0.0018 &      2.0 & 0.1 &   80 &   10 &  2.02 & $5\,618$ & 3.91 & active star                                                    \\ 
      7202957.02  &       - & 2 &  9 & 10.158 &    8.16912 &    0.00058 &   110.0257 &     0.0027 &      3.4 & 0.2 &  120 &   10 &  2.02 & $5\,618$ & 3.91 & id.                                                            \\
      8741367.01  &       - & 2 & 17 & 12.551 &    0.98907 &    0.00010 &    54.1067 &     0.0032 &      1.9 & 0.1 &  130 &   10 &     - &        - &    - & CoRoT 7b-like                                                  \\ 
      9044228.01  & 1988.01 & 1 & 14 & 14.145 &   0.934835 &   0.000002 &    65.5200 &     0.0007 &      1.3 & 0.1 &  240 &   10 &  1.66 & $4\,604$ & 4.03 & CoRoT 7b-like                                                  \\
      9475552.01  &       - & 2 & 16 & 14.975 &   0.843349 &   0.000002 &    64.9733 &     0.0010 &      1.4 & 0.1 &  300 &   10 &  1.11 & $4\,659$ & 4.31 & CoRoT 7b-like                                                  \\
      9729691.01  & 1751.01 & 1 & 14 & 14.548 &    8.69031 &    0.00094 &   110.1064 &     0.0045 &      2.4 & 0.1 & 1200 &  100 &  1.11 & $5\,052$ & 4.34 & active star, $P_\mathrm{rot} \sim 23$d; peak to peak 1\%       \\
      9729691.02  & 1751.02 & 1 & 18 & 14.548 &    20.9926 &     0.0013 &   107.1019 &     0.0033 &      2.7 & 0.2 & 1100 &  100 &  1.11 & $5\,052$ & 4.34 & id.                                                            \\
      9780149.01  &       - & 3 & 14 & 12.294 &    2.51729 &    0.00039 &    55.1147 &     0.0026 &      1.2 & 0.1 &  200 &   20 &     - &        - &    - &                                                                \\ 
      9872283.01  & 1815.01 & 3 & 12 & 13.884 &    3.12716 &    0.00011 &   107.3675 &     0.0014 &      1.6 & 0.1 &  630 &   40 &  1.80 & $4\,680$ & 3.98 &                                                                \\
      9909735.01  & 1779.01 & 1 & 17 & 13.297 &    4.66285 &    0.00011 &   109.0435 &     0.0010 &      3.0 & 0.1 & 1580 &   30 &  1.48 & $5\,666$ & 4.15 & active star, $P_\mathrm{rot} \sim 7.1$d; peak to peak 2\%      \\
      9909735.02  & 1779.02 & 1 & 13 & 13.297 &   11.81446 &    0.00079 &   125.1682 &     0.0023 &      2.7 & 0.1 &  850 &   50 &  1.48 & $5\,666$ & 4.15 & id.                                                            \\
      9965957.01  & 1911.01 & 2 &  9 & 15.196 &    5.75298 &    0.00054 &   110.0284 &     0.0054 &      1.7 & 0.2 &  780 &   90 &  0.87 & $5\,276$ & 4.55 &                                                                \\
      10190777.01 & 1937.01 & 3 & 17 & 13.590 &   1.411471 &   0.000035 &   106.5605 &     0.0011 &      1.5 & 0.1 &  380 &   20 &  0.70 & $4\,242$ & 4.55 &                                                                \\
      10332883.01 & 1880.01 & 3 & 17 & 14.440 &   1.151123 &   0.000040 &   105.3691 &     0.0016 &      2.0 & 0.1 &  470 &   30 &  0.69 & $3\,898$ & 4.49 &                                                                \\
      10464050.01 & 1851.01 & 3 & 17 & 14.769 &    4.46996 &    0.00022 &   109.5373 &     0.0017 &      1.4 & 0.1 &  590 &   60 &  0.82 & $5\,712$ & 4.63 &                                                                \\
      11187436.01 &       - & 3 & 13 & 15.625 &    5.90735 &    0.00005 &    66.9356 &     0.0003 &      1.2 & 0.1 & 5500 &  100 &  0.78 & $4\,813$ & 4.58 &                                                                \\ 
      11298298.01 & 1802.01 & 2 & 15 & 13.345 &    5.24518 &    0.00012 &   110.5189 &     0.0011 &      1.8 & 0.1 &  620 &   30 &  0.97 & $5\,868$ & 4.50 &                                                                \\
      11551692.01 & 1781.02 & 1 & 24 & 12.231 &   3.005148 &   0.000004 &    55.0771 &     0.0004 &      2.1 & 0.1 &  630 &   10 &  1.99 & $4\,720$ & 3.91 &                                                                \\
      11551692.02 & 1781.01 & 1 & 27 & 12.231 &   7.834483 &   0.000008 &    60.3851 &     0.0003 &      2.9 & 0.1 & 1900 &   20 &  1.99 & $4\,720$ & 3.91 &                                                                \\
      11551692.03 &       - & 1 &  8 & 12.231 &  58.017141 &   0.000030 &    84.8529 &     0.0012 &      4.6 & 0.1 & 1250 &   20 &  1.99 & $4\,720$ & 3.91 &                                                                \\
      11853878.01 & 1833.01 & 1 & 30 & 14.265 &    3.69302 &    0.00001 &    65.7330 &     0.0009 &      1.6 & 0.1 &  700 &   20 &  0.60 & $4\,300$ & 4.67 &                                                                \\ 
      11853878.02 &       - & 1 & 11 & 14.265 &    5.70959 &    0.00002 &    70.1174 &     0.0013 &      1.4 & 0.1 &  550 &   30 &  0.60 & $4\,300$ & 4.67 &                                                                \\
      11853878.03 &       - & 1 & 26 & 14.265 &    7.68426 &    0.00002 &    66.3617 &     0.0008 &      1.3 & 0.1 &  970 &   40 &  0.60 & $4\,300$ & 4.67 &                                                                \\
      11911561.01 &       - & 3 & 10 & 13.579 &   8.870498 &   0.000025 &    70.4779 &     0.0008 &      1.3 & 0.1 &  630 &   20 &     - &        - &    - &                                                                \\ 
      \hline
    \end{longtable}
  \end{landscape}
}


\longtab{4}{
\begin{longtable}{*{4}{c}*{2}{r@{$\pm$}l}*{2}{c}l}
  \caption{\label{table:fp}Parameters from the false positives.} \\
  \hline\hline
  KIC     & rank & SNR & Kp     & \multicolumn{2}{c}{period}              & \multicolumn{2}{c}{epoch}               & duration & depth & comments        \\
          &      &     & mag    & \multicolumn{2}{c}{[days]}              & \multicolumn{2}{c}{[HJD-$2\,454\,900$]} & [h]      & [ppm] &                 \\ 
  \hline
  \endfirsthead
  \caption{continued.}\\
  \hline\hline
  KIC     & rank & SNR & Kp     & \multicolumn{2}{c}{period}              & \multicolumn{2}{c}{epoch}               & duration & depth & comments        \\
          &      &     & mag    & \multicolumn{2}{c}{[days]}              & \multicolumn{2}{c}{[HJD-$2\,454\,900$]} & [h]      & [ppm] &                 \\ 
  \hline
  \endhead
  \hline
  \endfoot
     2853828  & 2 & 18 & 11.899 &   1.077470 &   0.000032 &   105.5274 &     0.0015 &      2.0 &   207 & pixel (2,8)                                  \\
     2997178  & 3 & 12 & 12.861 &    5.95079 &    0.00035 &   109.7210 &     0.0031 &      5.2 &   763 & pixel (2,5)                                  \\
     4570555  & 2 & 13 & 11.540 &     4.7494 &     0.0011 &    67.6398 &     0.0030 &      2.3 &   573 & pixel (3,6)                                  \\
     4666008  & 3 &  9 & 14.434 &    2.24795 &    0.00029 &   106.2166 &     0.0052 &      2.6 &   141 & pixel (6,2)                                  \\
     4913000  & 2 & 16 & 11.668 &    4.44222 &    0.00018 &   106.5901 &     0.0022 &      2.1 &   342 & pixel (1,4)                                  \\
     5302006  & 3 &  9 & 15.050 &     6.1442 &     0.0051 &     65.768 &      0.014 &      3.5 &  1718 & pixel (2,2)                                  \\
     5695904  & 3 &  9 & 15.586 &   0.424123 &   0.000093 &    65.1813 &     0.0040 &      2.0 &   274 & pixel (2,4)                                  \\
     5894073  & 1 & 11 & 13.748 &    10.5953 &     0.0057 &    69.9781 &     0.0068 &      5.1 &   511 & pixel (6,3)                                  \\
     9307105  & 3 & 12 & 15.023 &   13.54510 &    0.00052 &   108.9023 &     0.0024 &      4.4 &  1679 & pixel (1,4)                                  \\
     12691412 & 2 & 14 & 13.502 &     6.9030 &     0.0025 &    70.0241 &     0.0043 &      3.4 &   965 & pixel (7,4)                                  \\
     3836453  & 2 & 17 & 12.057 &   1.540567 &   0.000056 &   106.5835 &     0.0015 &      2.5 &   137 & contaminated by KIC~3836439                   \\
     3836522  & 3 & 15 & 14.844 &    1.54048 &    0.00010 &   106.5888 &     0.0034 &      2.3 &   573 & contaminated by KIC~3836439                   \\
     3836399  & 3 & 14 & 14.449 &    1.54019 &    0.00032 &    64.9981 &     0.0043 &      2.3 &   619 & contaminated by KIC~3836439                   \\
     2830919  & 1 &  8 & 14.753 &    1.54119 &    0.00044 &    65.0032 &     0.0049 &      2.1 &   445 & contaminated by KIC~3836439                   \\
     3325239  & 2 &  9 & 15.308 &    1.54114 &    0.00036 &    64.9837 &     0.0048 &      1.9 &   687 & contaminated by KIC~3836439                   \\
    \hline
  \end{longtable}
}

\longtab{5}{
\begin{longtable}{*{4}{c}*{2}{r@{$\pm$}l}*{2}{c}l}
  \caption{\label{table:fa}Parameters from the false alarms.} \\
  \hline\hline
  KIC     & rank & SNR & Kp     & \multicolumn{2}{c}{period}              & \multicolumn{2}{c}{epoch}               & duration & depth \\
          &      &     & mag    & \multicolumn{2}{c}{[days]}              & \multicolumn{2}{c}{[HJD-$2\,454\,900$]} & [h]      & [ppm] \\
  \hline
  \endfirsthead
  \caption{continued.}\\
  \hline\hline
  KIC     & rank & SNR & Kp     & \multicolumn{2}{c}{period}              & \multicolumn{2}{c}{epoch}               & duration & depth \\ 
          &      &     & mag    & \multicolumn{2}{c}{[days]}              & \multicolumn{2}{c}{[HJD-$2\,454\,900$]} & [h]      & [ppm] \\
  \hline
  \endhead
  \hline
  \endfoot
     3246908 & 3 &  8 & 14.756 &     3.8713 &     0.0017 &     66.168 &      0.036 &      3.4 &   443 \\
     3443220 & 2 &  8 & 13.615 &     7.0007 &     0.0048 &    71.1870 &     0.0094 &      3.9 &   472 \\
     3446775 & 2 &  9 & 10.949 &     4.9819 &     0.0036 &     66.168 &      0.011 &      4.3 &   219 \\
     3544602 & 2 & 10 & 15.011 &     4.4211 &     0.0016 &    65.8624 &     0.0040 &      2.4 &   519 \\
     3859729 & 3 &  9 & 13.852 &     6.4673 &     0.0018 &    65.4454 &     0.0044 &      1.9 &   759 \\
     3962913 & 2 & 10 & 13.729 &     1.8433 &     0.0040 &     64.851 &      0.039 &      2.2 &   176 \\
     3964162 & 1 &  8 & 15.331 &     4.1584 &     0.0015 &    65.2196 &     0.0064 &      2.0 &   838 \\
     4275305 & 3 &  8 & 12.573 &    0.84275 &    0.00037 &    65.3835 &     0.0082 &      1.8 &    83 \\
     4364467 & 2 & 10 & 14.653 &    11.8808 &     0.0084 &     70.875 &      0.014 &      1.8 &  1101 \\
     4366351 & 3 &  8 & 12.873 &    11.2557 &     0.0061 &    66.0898 &     0.0074 &      5.9 &   945 \\
     5015478 & 2 &  8 & 12.630 &     5.5103 &     0.0022 &    65.8242 &     0.0075 &      1.9 &   129 \\
     6043914 & 3 &  8 & 13.811 &    2.54285 &    0.00095 &    65.5311 &     0.0079 &      2.4 &   288 \\
     6305938 & 2 &  9 & 13.933 &     4.4399 &     0.0027 &     65.829 &      0.011 &      2.1 &   228 \\
     7946652 & 2 & 12 & 14.176 &     6.5196 &     0.0018 &    69.8381 &     0.0056 &      2.1 &   671 \\
     8241503 & 2 &  9 & 13.332 &     3.2996 &     0.0032 &     67.728 &      0.020 &      2.6 &   163 \\
     8625295 & 2 & 10 & 13.163 &    1.45826 &    0.00032 &    65.7306 &     0.0036 &      1.8 &   203 \\
     9827003 & 3 & 10 & 12.256 &    1.64217 &    0.00042 &    66.2096 &     0.0033 &      1.8 &   283 \\
    10743519 & 2 &  8 & 13.864 &     5.7054 &     0.0017 &    66.1704 &     0.0051 &      2.6 &   385 \\
    11047067 & 3 &  9 & 13.574 &     6.8768 &     0.0031 &    67.8136 &     0.0076 &      2.3 &   420 \\
    11350341 & 3 &  8 & 11.461 &     9.2973 &     0.0089 &     73.659 &      0.013 &      5.8 &   413 \\
    \hline
  \end{longtable}
}

\end{document}